\documentclass[%
 reprint,
 amsmath,amssymb,
 aps,
]{revtex4-2}

\usepackage{graphicx}
\usepackage{dcolumn}
\usepackage{bm}
\usepackage{tikz}
\usetikzlibrary{positioning, shapes.geometric}
\usepackage{braket}
\usepackage{mathrsfs}
\usepackage{amsmath}

\begin{document}

\preprint{APS/123-QED}

\title{Implementation of full-potential screened spherical wave based muffin-tin orbital for all-electron density functional theory}

\author{Aixia Zhang\textsuperscript{1}} 
\author{Qingyun Zhang\textsuperscript{1}} 
\email{zhangqy2@shanghaitech.edu.cn}
\author{Zhiyi Chen\textsuperscript{1}} 
\author{Yong Wu\textsuperscript{2}}
\author{Youqi Ke\textsuperscript{1}}
\email{keyq@shanghaitech.edu.cn}

\affiliation{\textsuperscript{1}School of Physical Science and Technology, ShanghaiTech University, Shanghai, 201210, China}
\affiliation{\textsuperscript{2}Institute of Applied Physics and Computational Mathematics, Beijing 100088, China}

\date{\today}

\begin{abstract}
Screened spherical wave (SSW) of the Hankel function features the complete, minimal and short-ranged basis set, presenting a compact representation for electronic systems. In this work,
we report the implementation of full-potential (FP) SSW based tight-binding linearized Muffin-Tin orbital (TB-LMTO) for all-electron density functional theory (DFT), and provide extensive tests on the robustness of FP-TB-LMTO and its high accuracy for first-principles material simulation. 
Through the introduction of double augmentation, SSW based MTO is accurately represented on the double grids including the full-space uniform and dense radial grids. Based on the the double augmentation, the accurate computation of full charge density, full potential,complex integral in the interstitial region and the total energy are all effectively addressed to realize the FP-TB-LMTO for DFT self-consistent calculations. By calculating the total energy,band structure, phase ordering, and elastic constants for a wide variety of materials, including normal metals, compounds, and diamond silicon, we domenstrate the highly accurate numerical implemetation of  FP-TB-LMTO for all-electron DFT in comparison with other well-established FP method.
The implementation of FP-TB-LMTO based DFT offers an important tool for the accurate first-principles tight-binding electronic structure calculations, particular important for the large-scale or strongly correlated materials.


\end{abstract}

\maketitle

\section{\label{sec:level1}Introduction}
Kohn-Sham density functional theory (KSDFT)~\cite{hohenberg1964density, kohn1965self}, widely used in condensed matter physics, materials science, and chemistry, maps the interacting many-body problem into a fictitious self-consistent single-particle problem. The Kohn-Sham equation is typically solved by expanding wavefunctions in a chosen basis set. 
The choice of basis functions forms the foundation of the algorithmic and numerical frameworks behind various first-principles DFT code packages developed over the years. The plane wave (PW)~\cite{RevModPhys.64.1045}, featuring infinite range and arbitrarily high accuracy, is a widely used and convenient choice, in combining with the development of the norm-conserving pseudopotential (NCPP) method\cite{PhysRevLett.43.1494}, projector augmented wave (PAW)\cite{blochl1994projector} method and the linearized augmented plane wave (LAPW) method \cite{Slater-1937} for different treatments of the effects of core electrons. The LAPW method\cite{FPLAPW1981,FPLAPW1984}, based on the muffin-tin (MT) partition of space, uses partial waves in the MT spheres and plane waves in interstitial region (that are smoothly matched at the MT sphere boundaries) to expand wavefunctions, and provides a highly accurate all-electron full-potential (FP) DFT electronic structure approach. However, PW extends all over the space resulting in a full Hamiltonian matrix, and typically requires around 100 basis functions per atom, presenting important limitation for large-scale material simulation. The reformulation of PW into spatially localized nonorthogonal generalized Wannier function extends to the linear-scaling materials modelling.~\cite{Skylaris2002NonorthogonalGW} The second-type of methods use the localized bases (not atom centered), including the real space grid based finite difference\cite{PhysRev.45.815,P1935IntroductionTQ,RevModPhys.72.1041,doi:10.1137/060651653} and finite element methods\cite{PhysRevB.52.5573}, wavelet method,\cite{10.1063/1.1768161} and etc. 
A third-type of basis-set approaches use localized atom-centered basis functions, including the methods of linear combination of atomic orbitals (LCAO)~\cite{LCAO}, Gaussian-type~\cite{1950RSPSA.200..542B} orbitals (GTO), and the spherical wave based muffin-tin orbitals (MTO), widely applied in molecular and condensed matter systems. The LCAO offers efficiency, accuracy and bonding insight, while it requires system-specific basis function tailoring. The GTO, popular in quantum chemistry, are efficient and enable simple algorithms but approximate wavefunctions poorly near nuclei due to their non-exponential form. 

Compared to other basis methods, the muffin-tin orbital method (MTO) methods, developed by O.K. Andersen and coworkers~\cite{Andersen1971}, features the complete, the minimal and high physical transparency~\cite{Zwierzycki_Andersen_2009,Methfessel1988ElasticCA}, has evolved to improve both computational efficiency and accuracy in electronic structure calculations. The MTOs are generally constructed by a spherical Hankel wave in the interstitial region and smoothly augmented by the partial waves inside the spheres. Both the Hankel function and partial waves are adapting to the muffin tin potential approximation, ensuring the minimal basis set of MTO. For the practical implementation of MTO, the linearization procedure is made to remove the energy dependence, and LMTO ultilizes the (multiple) bared spherical Hankel function of fixed kinetic energy $\kappa^2$ in the interstitial region, and partial waves at a set of energy center $\epsilon_{Rl}$ inside MT spheres. In the LMTO method, the eigenvalues and eigenvectors of the band calculations are correct to third and second order, respectively, in the deviation of the eigenvalue from the
 chosen energy center, while the LMTO itself is only correct to first order.\cite{PhysRevB.34.5253} However, the products of Hankel spherical wave are not naturally a sum of Hankel spherical waves, presenting the important difficulty for accurate representation of full charge density, full potential and the associated integrals in the interstitial region with complex topology. In past decades, great efforts have been spent to achieve the highly accurate full potential(FP) implementation of LMTO, including the construction of pseudo-LMTOs,\cite{Weyrich1988} interstitial interpolation technique\cite{nohara2016interpolation} and the method of smooth Hankel functions,\cite{bott1998nonsingular}. However, this first generation of LMTO faces the long-range problem of the bared Hankel functions, i.e., at $\kappa=0$, the bare s,p,d-LMTO waves falls off like the respective $1/r$, $1/r^2$, $1/r^3$ in the interstitial region, limiting its application to large scale simulations. To realize the short-ranged tight-binding LMTO, O.K.Andersen first introduced the screening technique to construct a highly localized envelop function by linear combination of bared Hankel functions with a set of screening factors $\alpha_{Rl}$, whithout changing the completeness.\cite{turek1997electronic, andersen1984explicit, Andersen1986PRB}
The TB-LMTO is often combined with the atomic sphere approximation (ASA)~\cite{andersen1984explicit,PhysRevB.34.2439,Andersen1986PRB}, which simplifies the potential by assuming spherical symmetry around each atom, presenting very high computational efficiency and good accuracy for close-packed structures, though it struggles with systems of lower symmetry. 
As the latest generation of MTOs, represented by the Exact MTO (EMTO)\cite{Andersen1995, structure2000physical, NMTO2012} and NMTO\cite{andersen2000muffin, NMTO2012} methods, marks a significant advancement in electronic structure calculations. These methods have been proven highly effective in simulating solid-state materials and nanoelectronic devices~\cite{KePRL2008, ZhangYanPRB, ChenZy2020, ChenZy2023, ZhangyuNEDCA}.
The EMTO method provides a general screeening approache by introducing hard screening spheres, eliminating the need to manually select the $\alpha_{Rl}$ parameters in TB-LMTO. Once the system's geometry and hard sphere radii are defined, a highly localized envelope function, called the screened spherical wave (SSW), is automatically generated.
The NMTO method further enhances the framework by linearly combining multiple (N-order) EMTOs at different energies, increasing both the flexibility and accuracy beyond the LMTO and providing a general scheme for generating the generalized Wannier functions. These compact MTO methods, featuring the short-ranged, the complete and minimal, are particularly useful for developing the order-N method for large-scale simulations and integrating with advanced techniques like the GW approximation~\cite{F_Aryasetiawan_1998} and dynamical mean-field theory (DMFT)~\cite{georges1996dynamical, anisimov1997first} for studying correlated electronic structures. 

However, despite these elegent feaures of SSW based MTOs, a highly accurate FP implementation of SSW based MTO is still lacking. In our recent work,\cite{AixiaZhangPRB} we presented a important step forward by demonstrating an accurate FP implementation of SSW based MTO for an all-electron DFT, which we generally call FP-TB-LMTO method. In this FP method, a general double augmentation scheme is introduced to accurately represent the TB-LMTOs, enabling the high accuracy for computing the full density, full potential, Hamiltonian integral, and total energy. In this work, we present the general algorithms and self-constained implementation details of the FP-TB-LMTO method, and provide an extensive tests on a variety of materials and discuss the robustness of TB-LMTO for the wide applicability. The rest of the paper is organized as follows:  Sec.\ref{formalism} presents the FP-TB-LMTO formalism for the self-consistent electronic structure calculation, including construction of  three component representation of basis function, charge density and potential and total energy calculation. In Sec.\ref{results}, we present numerical results and discussions to demonstrate the accuracy and applicability. Finally, we conclude our work in Sec.\ref{conclusion} and provide more information in Append.\ref{App.StoM}.

%
%
%
%

\section{Formalism}\label{formalism}
This section describes the basic ideas to realize the highly accurate full potential SSW based MTO method for self-consistent electronic structure calculations. In this paper,  we consider the MT geometry with nonoverlapping potential spheres with radii $s_R$. We use the atomic Rydberg units throughout this paper. In the following, we will first introduce  the construction of screened spherical wave, double augmentation for obtaining accurate three-component representation of TB-LMTO, full charge density and full potential, and then describe the calculation of  Hamiltonian, Overlap matrices and the calculation of total energy,and finally discuss the self-consistent procedures for implmentating FP-TB-LMTO based electronic structure calculations. 

\subsection{Construction of SSW based TB-LMTO}

To achieve the high-precision implementation, we here introduce double augmentation to form an accurate three-component representation of SSW based TB-LMTO $\chi^{\alpha}_{RL}$,  which includes the smooth SSW represented by a uniform real-space grid and the augmentation functions represented by a radial grid inside MT spheres.

\subsubsection{Envelope function: Screened Spherical Wave}
To construct highly localized TB-LMTO basis functions, SSWs $\Psi^{\alpha,I}_{RL}(\kappa^2,\bm r) $ are introduced as envelope functions centered on the site $R$, which are the solutions of Schrodinger equation in the interstitial region, namely 
\begin{equation}
\label{SSW-equation}
\{\nabla^2 + \kappa^2 \} \Psi^{\alpha,I}_{RL} = 0
\end{equation}
where $L\equiv lm$ ($l\le l_{max}$, $(l_{max}+1)^2$ gives the number of MTO on R), $\alpha$ denotes the screening representation and kinetic energy $\kappa^2 = \epsilon -V_{MTZ}$ (where $V_{MTZ}$ is the constant potential in interstitial region in MT potential approximation). To make $\Psi^{\alpha,I}_{RL}$  short-range or screened in real space, a set of non-overlapping screening spheres with radius $a_{Rl}$($\le s_{R}$ which is the potential sphere radius.) are introduced, and then boundary conditions are imposed on SSWs $\Psi^{\alpha,I}_{RL}$  so that it equals pure spherical harmonics on its own $a$-sphere and vanishes on the other $a$-spheres, for $l'\le l_{max}$, namely
\begin{equation} \label{eq:Projection}
    \mathscr{\hat{P}}_{R'L'}(a_{R'l'}) \Psi^{\alpha,I}_{RL}(\bm r) = \delta_{RR'}\delta_{LL'}
\end{equation}
where  $\mathscr{\hat{P}}_{R'L'}$ denotes the projection operator. To proceed,  SSWs in interstitial region can be expressed as superposition of the bared solutions of wave equation Eq.~(\ref{SSW-equation}), namely the spherical Hankel functions,
\begin{equation}\label{eq:M-matrix}
    \begin{aligned}
      \Psi^{\alpha,I}_{RL}(\kappa^2,\bm r)
    & =  \sum_{R'L'} H_{L'}(\kappa^2, \bm r_{R'}) M^{a}_{R'L',RL}(\kappa^2),
    \end{aligned}
\end{equation}
where $\kappa^2$ can be both positive and negtive values, $l'\le l_{max}$, and  $H_{L'}(\kappa^2, \bm r_{R'})\equiv -i\kappa^{l'+1}  h_{l'}^{(1)}(\kappa r)  Y_{L'}(\hat{\bm r}_{R'})$, $h^{(1)}(\kappa r) $ is the spherical Hankel function of the first kind.\cite{nohara2016interpolation} 
Here, $M^{a}$ is a highly sparse matrix, can be solved by applying the boundary condition of  $\Psi^{\alpha,I}_{RL}$~\cite{nohara2016interpolation}, please refer to Appendix \ref{App.StoM} for details. 

For the convenience of subsequent augmentations, the SSW $\Psi^{\alpha,I}_{RL}(\kappa^2,\bm r)$ can be rewritten with a single-center expansion form, equivalent to Eq.~(\ref{eq:M-matrix}), namely be expanded in real spherical harmonics $Y_{L'}(\hat{\bm r}_{R'})$ around a site $R'$,
\begin{equation}
    \label{SSW}
    \begin{aligned}
      \Psi^{\alpha,I}_{RL}& (\kappa^2,\bm r)= f_{Rl}^a(\kappa^2, r_R)Y_L(\hat{\bm r}_R)\delta_{RR'}\delta_{LL'} \\
    & + \sum_{L'}g_{R'l'}^a(\kappa^2,r_{R'})Y_{L'}(\hat{\bm r}_{R'})S_{R'L'RL}^a(\kappa^2),
    \end{aligned}
\end{equation}
where $l'\le l_{high}$ which is chosen large enough to ensure the convergence. The sparse slope matrix $S_{R'L'RL}^a(\kappa^2)$ is connected to matrix $M^a$ (see Appendix \ref{App.StoM}). Here the head function $f_{Rl}^a$  and tail  function $g_{Rl}^a$ are defined as the linear combination of the spherical Bessel $j$ and Neumann $n$ functions, namely
   \begin{equation}
    \begin{aligned}
    f_{Rl}^{a}(\kappa^2, r_{R}) & \equiv A_{Jf_{ Rl}}j_{l}(\kappa^2,r_{ R}) + B_{Nf_{Rl}}n_{l}(\kappa^2,r_{R}), \\
    g_{Rl}^{a}(\kappa^2, r_{R}) & \equiv  A_{Jg_{ Rl}}j_{l}(\kappa^2,r_{R}) + B_{Ng_{Rl}}n_{l}(\kappa^2,r_{R}),
    \end{aligned} \label{fgjn}
    \end{equation}
where the coefficients $A_{Jf_{ Rl}}$, $B_{Nf_{Rl}}$, $A_{Jg_{Rl}}$ and $B_{Ng_{Rl}}$ can be determined by setting the boundary conditions of $f_{Rl}^{a}$ and $g_{Rl}^{a}$ as follows. 
 By applying the projection $\mathscr{\hat{P}}_{R'L'}(r)$ on Eq.~(\ref{SSW}), then we obtain
\begin{equation}
    \begin{aligned}
         \mathscr{\hat{P}}_{R'L'}(\bm r_{R'})\Psi^{\alpha,I}_{RL}&(\kappa^2,\bm r_R) =f_{Rl}^a(\kappa^2, r_{R}) \delta_{RR'}\delta_{LL'}  \\
         & + g_{R'l'}^a(\kappa^2,r_{R}) S_{R'L'RL}^a(\kappa^2).
    \end{aligned}\label{ProjPhi}
\end{equation}
To satisfy the boundary condition of SSW in Eq.~(\ref{eq:Projection}), one can fix the head $f_{Rl}^{a}$ and tail $g_{Rl}^{a}$ functions at screening $a$-spheres for all active channels ($l\le l_{max}$) to the boundary values (other choices are possible, here we take the convention of Ref.~\cite{Vitos2007})
\begin{equation}
    \label{boundary cond.}
    \begin{aligned}
        f_{Rl}^a(\kappa^2,a_R)=1,\   &  \  g_{Rl}^a(\kappa^2, a_R)=0; \\
        \frac{\partial f_{Rl}^a(\kappa^2, r)}{\partial r}|_{a_R}=0,\   &  \ \frac{\partial  g_{Rl}^a(\kappa^2, r)}{\partial r}|_{a_R}=\frac{1}{a_R}.
    \end{aligned}
\end{equation}
For all in-active channels, namely $l'>l_{max}$ in Eq.~(\ref{SSW}), the tail functions $g_{Rl}^a$  are setted to Bessel function. In such a way, the constructed SSWs features the important localization inside the interstitial region, which is desirable for many applications. 


\begin{figure*}[htbp]
    \centering
    \includegraphics[scale=0.900]{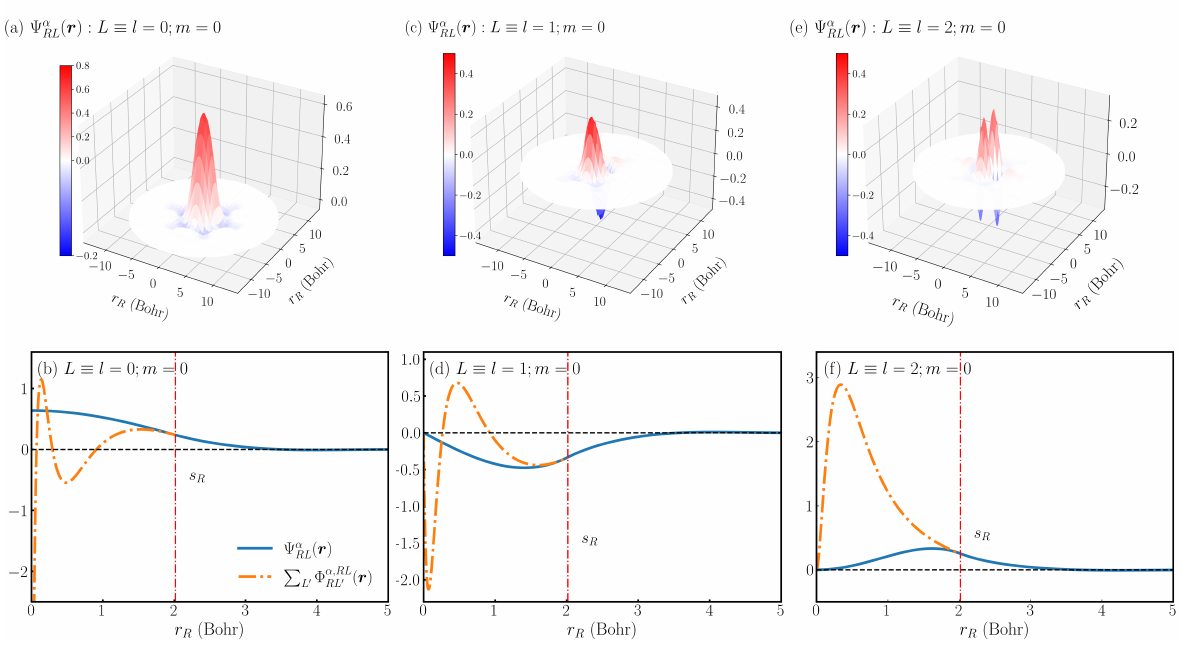} %
        \caption{
        Basis functions in real space for the Ni FCC structure.  \textbf{3D surface} plots of the Smooth SSW $\Psi^\alpha_{RL}(\bm{r})$ for 
        \textbf{(a)} $L \equiv l = 0, m = 0$; 
        \textbf{(c)} $L \equiv l = 1, m = 0$; 
        \textbf{(e)} $L \equiv l = 2, m = 0$, 
        where the $z$ axis represents the magnitude of $\Psi^\alpha_{RL}(\bm{r})$, and $r_R$ represents the distance between the grid point $\bm{r}$ and atomic site of $R$. 
        \textbf{2D diagrams} for
        \textbf{(b)} $L \equiv l = 0, m = 0$; 
        \textbf{(d)} $L \equiv l = 1, m = 0$; 
        \textbf{(f)} $L \equiv l = 2, m = 0$, with the vector direction $\bm{r} = \frac{1}{\sqrt{6}}[x, y, -2z]$ within the (111) plane.
        The parameters used in the construction of the SSW are:  average Wigner-Seitz cell radius (WSA) $\omega$ =  2.52 Bohr, $a_R = 0.7$ $\omega$ = 1.764 Bohr, $s_R = 0.8$ $\omega$ = 2.016 Bohr, and $\kappa^2 = 0.0$~Ry.
    }\label{Fig:Basis-Func111} 
\end{figure*}

\subsubsection{Auxiliary Augmentation}
 Due to the facts that the product of SSWs is not naturally SSWs and the topology of interstitial region is complex, it is of great difficulty for accurately representing the charge density and solving for the full potential.
 To tackle this problem for realizing highly accurate full potential SSW based TB-LMTO calculation, we introduce an auxiliary augmentation to represent the SSWs, 
\begin{equation}\label{eq:auxiliarySSW1}
 \Psi^{\alpha}_{RL}(\bm r) = \Psi^{\alpha,I}_{RL}(\bm r) + \sum_{R'L'} \tilde{\Psi}^{\alpha,RL}_{R'L'}(\bm r_{R'}),
\end{equation}
where the augmentation functions $\tilde{\Psi}^{\alpha,RL}_{R'L'}(\bm r)$ inside the muffin-tin spheres are slowly-varying function and smoothly connected to the SSWs at the boudnary of augmentation spheres $s_R$. Here, $\Psi^{\alpha}_{RL}(\bm r)$ is defined as smoothed SSW, which extends into the MT spheres and is slowly-varying function in the full space. The auxiliary augmentation functions, $\tilde{\Psi}^{\alpha,RL}_{R'L'}$ in Eq.~(\ref{eq:auxiliarySSW}), are constructed to ensure continuity and differentiability with $\Psi^{\alpha,I}_{RL}$ at the boundary $s_{R'}$ by matching the values and slopes of $f_{Rl}^{a}$ and $g_{Rl}^{a}$ in Eq.~(\ref{SSW}) on $s_{R}$ with auxiliary smooth functions $\tilde{f}_{Rl}^{a}$ and $\tilde{g}_{Rl}^{a}$,
   \begin{equation}\label{augauxiliary}
    \begin{aligned}
    \tilde{f}_{Rl}^{a}( r_{R}) &= A_{\tilde{\psi}f_{ Rl}}\tilde{\psi}_{l}(\beta_1,r_{ R}) + B_{\tilde{\psi}f_{Rl}}{\tilde{\psi}}_{l}(\beta_2,r_{R}), \\
    \tilde{g}_{Rl}^{a}(r_{R}) & = A_{\tilde{\psi}g_{ Rl}}\tilde{\psi}_{l}(\beta_1,r_{R}) + B_{\tilde{\psi}g_{Rl}}{\tilde{\psi}}_{l}(\beta_2,r_{R})
    \end{aligned}
    \end{equation}
    where $A_{\tilde{\psi}f_{ Rl}}$, $A_{\tilde{\psi}g_{ Rl}}$, $B_{\tilde{\psi}f_{ Rl}}$, $B_{\tilde{\psi}g_{ Rl}}$ are coefficients determined by the boundary conditions. Here ,we adopt $\tilde{\phi}_l$ in the form $\tilde{\psi}_{l}(\beta,r) = r^l e^{-\beta^2 r^2}$, where $\beta$ is a tunable parameter for ensuring the smoothness inside MT spheres. It should be mentioned that 
$\tilde{f}_{Rl}^{a}$, $\tilde{g}_{Rl}^{a}$ can be constructed to smoothly connect with SSWs to an arbitrary order. As a result, we can write $ \tilde{\Psi}^{\alpha,RL}_{R'L'}(\bm r)$ in the form, for $r_R \le s_R$,
    \begin{equation}\label{smotthPSi-Define}
    \begin{aligned}
         \tilde{\Psi}^{\alpha,RL}_{R'L'}(\bm r) &=  \tilde{f}_{Rl}^a(r_R)Y_L(\bm {\hat{r}}_R)\delta_{RR'}\delta_{LL'} \\
         & +  \tilde{g}_{R'l'}^a(r_{R'})Y_{L'}(\hat{\bm r}_{R'})S_{R'L'RL}^a(\kappa^2).
    \end{aligned}
    \end{equation}
 Due to the facts that the slowly-varying $\Psi^{\alpha}_{RL}(\bm r)$ and $\tilde{\Psi}^{\alpha,RL}_{R'L'}(\bm r)$ can be represented accurately on the respective real-space uniform grid and radial grid inside MT spheres, we can reach a high-precision representation for the SSW, namely
\begin{equation}
\label{eq:auxiliarySSW}
\Psi^{\alpha,I}_{RL}(\bm r) = \Psi^{\alpha}_{RL}(\bm r) - \sum_{R'L'} \tilde{\Psi}^{\alpha,RL}_{R'L'}(\bm r_{R'}).
\end{equation}

\subsubsection{Partial-wave Augmentation}

 To form a TB-LMTO $\chi_{RL}$ centered at a site $R$, the second augmentation with partial wave and its derivative is introduced,\cite{Andersen1971,andersen1984explicit,Andersen1995,structure2000physical,andersen1971electronic,andersen2000muffin,NMTO2012} 
\begin{equation}\label{eq:genASSW}
\begin{aligned}
   \chi^{\alpha}_{RL}(\epsilon, \bm r)=  \Psi^{\alpha,I}_{RL}(\kappa^2,\bm r) + \sum_{R'L'} \Phi^{\alpha,RL}_{R'L'}(\epsilon_{l'},\bm r_{R'}).
\end{aligned}
\end{equation}
in which, inside the muffin-tin potential sphere, the augmentation function $ \Phi^{\alpha,RL}_{R'L'}(\epsilon_{l'},\bm r_R') $ are chosen as the linear combination of partial wave $\phi_{Rl}(\epsilon_{l'}, r)$ and its energy derivatives $\dot{\phi}_{Rl}(\epsilon_{l'}, r)$, which are constructed to ensure the continuity and differentiability with $\Psi^{\alpha,I}_{RL}$ at the sphere boundary $s_{R'}$. For each spheres, we can define ${f}_{Rl}^{a,\phi}$ and ${g}_{Rl}^{a,\phi}$  for $r_R<s_R$,  
   \begin{equation}\label{augpartitial}
    \begin{aligned}
    f_{Rl}^{a,\phi}(\epsilon_l, r_{R}) & \equiv A_{\phi f_{ Rl}}\phi_{l}(\epsilon_l,r_{ R}) + B_{\phi f_{Rl}}\dot{\phi}_{l}(\epsilon_l,r_{R}) \\
    g_{Rl}^{a,\phi}(\epsilon_l, r_{R}) & \equiv  A_{\phi g_{ Rl}}\phi_{l}(\epsilon_l,r_{R}) + B_{\phi g_{Rl}}\dot{\phi}_{l}(\epsilon_l,r_{R})
    \end{aligned}
    \end{equation}
    where $A_{\phi f_{ Rl}}$, $A_{\phi g_{ Rl}}$, $B_{\phi f_{Rl}}$, $B_{\phi g_{Rl}}$ are coefficients determined by matching the value and derivative of $f_{Rl}^{a}$ and $g_{Rl}^{a}$ in the  Eq.~(\ref{SSW}) at $s_R$. 
     As a result, $ {\Phi}^{\alpha,RL}_{R'L'}(\epsilon_{l'},\bm r_{R'})$ for $\bm r$ inside the MT spheres can be explicitly written as,
    \begin{equation} \label{Phi-Define}
    \begin{aligned}
         {\Phi}^{\alpha,RL}_{R'L'}(\epsilon_{l'}&,\bm r_{R'})=   f_{Rl}^{a,\phi}(\epsilon_l, r_R)Y_L(\hat{\bm {r}}_R)\delta_{RR'}\delta_{LL'} \\
         & +  {g}_{R'l'}^{a,\phi}(\epsilon_{l'}, r_{R'})Y_{L'}(\hat{\bm r}_{R'})S_{R'L'RL}^a(\kappa^2),    
    \end{aligned}
    \end{equation}
which can be described accurately on a radial grid inside each MT sphere. 

\subsubsection{Three-Component Representation of TB-LMTO}
By combining   Eq.~(\ref{eq:auxiliarySSW}) and Eq.~(\ref{eq:genASSW}), the TB-LMTO $\chi_{RL}(\bm r)$ in Eq.~(\ref{eq:genASSW}) can be rewritten in a three-component form,
\begin{equation}\label{eq:ASSW} 
  \chi_{RL}(\bm r) = \Psi^{\alpha}_{RL}(\bm r) + \sum_{R'L'} \{\Phi^{\alpha,RL}_{R'L'}(\bm r) - \tilde{\Psi}_{R'L'}^{\alpha,RL}(\bm r) \}, 
\end{equation}
where $\Psi^{\alpha}_{RL}(\bm r)$ is smooth over all space, and $\Phi^{\alpha,RL}_{R'L'}(\bm r)$ is a fast-varying function inside MT sphere while $\tilde{\Psi}_{R'L'}^{\alpha,RL}(\bm r)$ is smooth.  Therefore, two different grid systems can be introduced to accurately represent the different part of SSW based TB-LMTO, namely using a uniform sparse grid for  $\Psi^{\alpha}_{RL}(\bm r)$ and a dense radial grid for functions inside the MT sphere, to enable a practical and accurate implementation of FP-TB-LMTO method. To better understand the properties of the SSW based TB-LMTO by construction, Fig.~\ref{Fig:Basis-Func111} (a), (c) and (e) show the $\Psi^{\alpha}_{RL}(\bm r)$ for the Ni FCC on the (111) plane and Fig.~\ref{Fig:Basis-Func111}~ (b), (d) and (f) show the functions $\Psi^{\alpha}_{RL}(\bm r)$ and  $\sum_{R'L'} \Phi^{\alpha,RL}_{R'L'}(\bm r)$ along the direction $\bm{r} = \frac{1}{\sqrt{6}}[x, y, -2z]$,  for the respective $l$= 0, 1 and 2 with fixed $m=0$ (the system paramters as shown in the caption). The distance between the nearest-neighbor sites in this system is 4.56 Bohr. It can be seen that $\Psi^\alpha_{RL}(\bm{r})$ is smooth and highly localized, the augmentation $\sum_{R'L'} \Phi^{\alpha,RL}_{R'L'}(\bm r)$ change rapidly inside the MT spheres. In the 3D surface plots, it is clear that SSW in the interstitial region quickly decays to zero (in white) for all calculated $lm$ in the plane. As shown in Fig.~\ref{Fig:Basis-Func111}(b)(d)(f), it is evident that,  $\Psi^{\alpha}_{RL}(\bm r)$ in (d) decays to almost zero at $r_R$= 2.5, 3.5 and 3.4 Bohr  (outside the MT spheres) for the respective $l$= 0, 1, 2 orbitals along the direction, presenting strong localization behavior of SSW. 
%

After the introduction of three-component representation of the TB-LMTO in Eq.~(\ref{eq:ASSW}), the product of two $\chi$ can be also written in a three-component form as follows
\begin{equation}\label{eq:bi-linear} 
\begin{aligned}
    \chi_{RL}(\bm r)\chi_{R'L'} &(\bm r) =\Psi^{\alpha}_{RL}(\bm r)\Psi^{\alpha}_{R'L'}(\bm r) + \sum_{R_1L_1L_2} \{\Phi^{\alpha,RL}_{R_1L_1}(\bm r) \\
   &   \Phi^{\alpha,R'L'}_{R_1L_2}(\bm r) - \tilde{\Psi}_{R_1L_1}^{\alpha,RL}(\bm r) \tilde{\Psi}_{R_1L_2}^{\alpha,R'L'}(\bm r) \},
\end{aligned}
\end{equation}
which can provide an accurate method for calculating the charge density and various integrations in the following.

\subsection{Overlap Matrix}

With Eqs.~(\ref{eq:ASSW}, \ref{eq:bi-linear}), the overlap matrix elements $O_{R'L',RL} = \langle \chi_{R'L'}| \chi_{RL} \rangle$ can be written into a three-component form,
\begin{equation}
\label{O-Matrix-elements}
\begin{aligned}
 O_{R'L',RL}  &=  \langle \Psi^{\alpha}_{R'L'}| \Psi^{\alpha}_{RL} \rangle + \sum_{  R_1L_1}\{ \langle { \Phi^{\alpha,R'L'}_{R_1L_1} } | {\Phi^{\alpha,RL}_{R_1L_1}} \rangle \\
 &\quad  - \langle{\tilde{\Psi}^{\alpha,R'L'}_{R_1L_1} } | {\tilde{\Psi}^{\alpha,RL}_{R_1L_1} }  \rangle \} .
\end{aligned}
\end{equation}
where the first term can be calculated by an accurate uniform-grid integration, and the other two terms are calculated on the radial grid inside MT spheres. It should be noted that, the third term, namely  $\sum_{R_1L_1}\langle{\tilde{\Psi}^{\alpha,R'L'}_{R_1L_1} } | \tilde{\Psi}^{\alpha,RL}_{R_1L_1}\rangle$ fully cancel out the integration of the first term inside MT spheres, presenting a contribution of interstitial intergration to the overlap matrix. The overlap integration of partial-wave augmentation inside MT spheres can be explicitly reformulted as 
\begin{equation}
\label{O1R-Matrix-elements}
\begin{aligned}
 \langle { \Phi^{\alpha,R'L'}_{R_1L_1} } &| {\Phi^{\alpha,RL}_{R_1L_1}} \rangle = \langle { f^{a,\phi}_{R_1l_1} | f^{a,\phi}_{R_1l_1} } \rangle \delta_{RR_1}\delta_{R'R_1} \delta_{LL_1} \delta_{L'L_1} \\
& + \langle { g^{a,\phi}_{R_1l_1} | f^{a,\phi}_{R_1l_1} } \rangle \delta_{RR_1} \delta_{LL_1} S_{R_1L_1R'L'}^a \\
& + \langle { f^{a,\phi}_{R_1l_1} | g^{a,\phi}_{R_1l_1} } \rangle \delta_{R'R_1} \delta_{L'L_1} S_{R_1L_1RL}^a  \\
& + \langle { g^{a,\phi}_{R_1l_1} | g^{a,\phi}_{R_1l_1} } \rangle S_{R_1L_1R'L'}^a S_{R_1L_1RL}^a ,
\end{aligned}
\end{equation}
where the factors  $\langle { f^{a,\phi}_{R_1l_1} | f^{a,\phi}_{R_1l_1} } \rangle$, $\langle { f^{a,\phi}_{R_1l_1} | g^{a,\phi}_{R_1l_1} } \rangle$, $\langle { g^{a,\phi}_{R_1l_1} | f^{a,\phi}_{R_1l_1} } \rangle$ and $\langle { g^{a,\phi}_{R_1l_1} | g^{a,\phi}_{R_1l_1} } \rangle$ are the radial integration of the products of $f^{a,\phi}_{R_1l_1}$ and $g^{a,\phi}_{R_1l_1}$ radial functions defined in Eq.~(\ref{augpartitial}), for example $\langle { f^{a,\phi}_{Rl} | f^{a,\phi}_{Rl} }  \rangle \equiv \int_0^{s_{R}} r_{R}^2dr_{R} f^{a,\phi *}_{Rl} (r_{R}) f^{a,\phi}_{Rl}(r_{R})$. 
The calculation of $\langle{\tilde{\Psi}^{\alpha,R'L'}_{R_1L_1} } | {\tilde{\Psi}^{\alpha,RL}_{R_1L_1} }  \rangle$ is similar to the calculation of $\langle { \Phi^{\alpha,R'L'}_{R_1L_1} } | {\Phi^{\alpha,RL}_{R_1L_1}} \rangle$,by using $ \tilde{f}^{a}$ and $ \tilde{g}^{a}$ defined in Eq.~(\ref{augauxiliary}) in the place of $f^{a,\phi}$ and $g^{a,\phi}$.

\subsection{Hamiltonian Matrix}
The Hamiltonian matrix contains the contributions of kinetic energy and full potential, namely, $H_{R'L',RL}=  T_{R'L',RL} +  V_{R'L',RL}$. The full effective potential $V(\bm r)$  is also in the three-component form  (as shown in Sec.\ref{Sec-EffV}), 
\begin{equation}\label{eq:potential} 
   V(\bm r)=\tilde{V}^0(\bm r)+\sum_{R} \{V^{1}_{R}(\bm r_{R})-\tilde{V}^{2}_{R}(\bm r_{R}) \}, 
\end{equation}
with $V^1_{R}(\bm r) \equiv \sum_L V^1_{RL}(r) Y_{L}(\hat{\bm r}_R)$ and $\tilde{V}^2_{R}(\bm r)  \equiv  \sum_L \tilde{V}^2_{RL}(r) Y_{L}(\hat{\bm r}_R)$.  $\tilde{V}_0$ represents smooth effective potential, which is extending through the unit cell, tabulated on a real-space mesh. $V^1_{R}$ represents the actual  potential inside the sphere $s_R$. $\tilde{V}^2_{R}$ cancels out the $\tilde{V}^0$ within $s_R$, presenting an accurate representation of interstital potential. 
By using the Eqs.~(\ref{eq:ASSW},\ref{eq:bi-linear},\ref{eq:potential}),  the full potential matrix elements $ V_{R'L',RL}  = \bra{\chi_{R'L'}}  \hat{V}   \ket{\chi_{RL}}$ can be calculated in three parts as follows, 
\begin{equation}
\label{V-Matrix-elements}
\begin{aligned}
V_{R'L',RL} & =  \bra{ \Psi^{\alpha}_{R'L'}}\tilde{V}^0 \ket{\Psi^{\alpha}_{RL}} \\
& + \sum_{  R_1L_1L_2L_3}\{ \bra{ \Phi^{\alpha,R'L'}_{R_1L_2} } V^1_{R_1L_3} \ket{\Phi^{\alpha,RL}_{R_1L_1}}  \\
 &  - \bra{\tilde{\Psi}^{\alpha,R'L'}_{R_1L_2} } \tilde{V}^2_{R_1L_3} \ket{\tilde{\Psi}^{\alpha,RL}_{R_1L_1} } \}, 
\end{aligned}
\end{equation}
with the first term calculated by the numerical integration on a real-space uniform grid and  the others are integrated on the radial grid inside MT spheres. Inside MT spheres, the contribution can be reformulated as
\begin{equation}
\label{V-radial}
\begin{aligned}
 \bra{ \Phi^{\alpha,R'L'}_{R_1L_2} } & V^1_{R_1L_3} \ket{\Phi^{\alpha,RL}_{R_1L_1}}  = C_{L_1L_2L_3}\\
\times & \{ \langle { f^{a,\phi}_{R_1l_2} | V^1_{R_1L_3}| f^{a,\phi}_{R_1l_1} } \rangle \delta_{RR_1}\delta_{R'R_1} \delta_{LL_1} \delta_{L'L_2} \\
& + \langle { g^{a,\phi}_{R_1l_2} | V^1_{R_1L_3}| f^{a,\phi}_{R_1l_1} } \rangle \delta_{RR_1} \delta_{LL_1} S_{R_1L_1R'L'}^a \\
& + \langle { f^{a,\phi}_{R_1l_2} | V^1_{R_1L_3}| g^{a,\phi}_{R_1l_1} } \rangle \delta_{R'R_1} \delta_{L'L_1} S_{R_1L_1RL}^a  \\
& + \langle { g^{a,\phi}_{R_1l_2} | V^1_{R_1L_3}| g^{a,\phi}_{R_1l_1} } \rangle S_{R_1L_1R'L'}^a S_{R_1L_1RL}^a \},
\end{aligned}
\end{equation}
where $C_{L_1L_2L_3}$ denotes the Clebsch–Gordan coefficients, and the factors $\langle { f^{a,\phi}_{R_1l_2} | V^1_{R_1L_3}| f^{a,\phi}_{R_1l_1} }\rangle$, $ \langle { g^{a,\phi}_{R_1l_2} | V^1_{R_1L_3}| f^{a,\phi}_{R_1l_1} } \rangle$, $\langle { f^{a,\phi}_{R_1l_2} | V^1_{R_1L_3}| g^{a,\phi}_{R_1l_1} } \rangle$ and $\langle { g^{a,\phi}_{R_1l_2} | V^1_{R_1L_3}| g^{a,\phi}_{R_1l_1} } \rangle$ are the radial integrations, for example
    \begin{equation}
    \begin{aligned}
         \langle { f^{a,\phi}_{Rl} | V^1_{RL'}| f^{a,\phi}_{Rl} }  \rangle = \int_0^{s_{R}} r_{R}^2 f^{a,\phi *}_{Rl} (r_{R}) V^1_{RL'}(r_R) f^{a,\phi}_{Rl}(r_{R})dr_R. 
    \end{aligned}
    \end{equation}   
The calculation of $\langle{\tilde{\Psi}^{\alpha,R'L'}_{R_1L_2} } |\tilde{V}^2_{R_1L_3} | {\tilde{\Psi}^{\alpha,RL}_{R_1L_1} }  \rangle$ in Eq.~(\ref{V-Matrix-elements}) is similar to $\langle { \Phi^{\alpha,R'L'}_{R_1L_2} } |{V}^1_{R_1L_3}| {\Phi^{\alpha,RL}_{R_1L_1}} \rangle$.

For the elements of kinetic energy operator  $T_{R'L',RL} = \bra{\chi_{R'L'}} -\nabla^2 \ket{\chi_{RL}}$, we can write, 
\begin{equation}
\label{T-Matrix-elements}
\begin{aligned}
T_{R'L',RL} & =  \bra{ \Psi^{\alpha}_{R'L'}} -\nabla^2  \ket{\Psi^{\alpha}_{RL}} \\
& + \sum_{  R_1L_1L_2}\{ \bra{ \Phi^{\alpha,R'L'}_{R_1L_2} } -\nabla^2  \ket{\Phi^{\alpha,RL}_{R_1L_1}}  \\
 &  - \bra{\tilde{\Psi}^{\alpha,R'L'}_{R_1L_2} } -\nabla^2  \ket{\tilde{\Psi}^{\alpha,RL}_{R_1L_1} } \}.
\end{aligned}
\end{equation}
by utilizing the relation of Eq.~(\ref{SSW-equation}), we can obtain
\begin{equation}
\label{T-Matrix-elements}
\begin{aligned}
T_{R'L',RL}  = & \kappa^2 \langle \Psi^{\alpha,I}_{R'L'} | \Psi^{\alpha,I}_{RL} \rangle \\
&+\sum_{  R_1L_1L_2} \bra{ \Phi^{\alpha,R'L'}_{R_1L_2} }  -\nabla^2 \ket{\Phi^{\alpha,RL}_{R_1L_1}},  
\end{aligned}
\end{equation}
in which the first term is the interstitial contribution connected to the interstitial overlap matrix, and the second term is the actual contribution inside the MT spheres. Here,
\begin{equation}
\begin{aligned}
\langle \Psi^{\alpha,I}_{R'L'} | \Psi^{\alpha,I}_{RL} \rangle =\langle \Psi^{\alpha}_{R'L'} | \Psi^{\alpha}_{RL} \rangle - \sum_{  R_1L_1} \langle{\tilde{ \Psi}^{\alpha,R'L'}_{R_1L_1} }|  {\tilde{ \Psi}^{\alpha,RL}_{R_1L_1}}\rangle 
\end{aligned}
\end{equation}
avoiding the numerical problems in the gradients of  $\Psi^{\alpha}_{RL}$ due to the truncation for $l\le l_{high}$ in the augmentation. For the kinetic energy integration inside MT spheres, we can rewrite it as
\begin{equation}
\begin{aligned}
& \bra{ \Phi^{\alpha,R'L'}_{R_1L_1} }  -\nabla^2 \ket{\Phi^{\alpha,RL}_{R_1L_1}}  \\
& = -\langle { f^{a,\phi}_{R_1l_1} |\Delta_r| f^{a,\phi}_{R_1l_1} } \rangle \delta_{RR_1}\delta_{R'R_1} \delta_{LL_1} \delta_{L'L_1} \\
&  - \langle { g^{a,\phi}_{R_1l_1} |\Delta_r| f^{a,\phi}_{R_1l_1} } \rangle \delta_{RR_1} \delta_{LL_1} S_{R_1L_1R'L'}^a \\
& - \langle { f^{a,\phi}_{R_1l_1} |\Delta_r| g^{a,\phi}_{R_1l_1} } \rangle \delta_{R'R_1} \delta_{L'L_1} S_{R_1L_1RL}^a  \\
& -\langle { g^{a,\phi}_{R_1l_1} |\Delta_r| g^{a,\phi}_{R_1l_1} } \rangle S_{R_1L_1R'L'}^a S_{R_1L_1RL}^a, 
\end{aligned}
\end{equation}
where the prefactors $\langle { f^{a,\phi}_{R_1l_1} |\Delta_r| f^{a,\phi}_{R_1l_1} } \rangle$, $\langle { g^{a,\phi}_{R_1l_1} |\Delta_r| f^{a,\phi}_{R_1l_1} } \rangle$, $\langle { f^{a,\phi}_{R_1l_1} |\Delta_r| g^{a,\phi}_{R_1l_1} } \rangle $, $\langle { g^{a,\phi}_{R_1l_1} |\Delta_r| g^{a,\phi}_{R_1l_1} } \rangle $ are the radial integrals, with $\Delta_r=-\frac{1}{r^2}\frac{\partial}{\partial r}(r^2\frac{\partial}{\partial r}) + \frac{l(l+1)}{r^2}$.
Based on the uniform grid in full space and radial grid inside muffin-tin spheres, the Hamiltonian and Overlap matrices of TB-LMTO can be accurately calculated to proceed to solve the Kohn-Sham equation to obtain the full charge density.

\subsection{Electron Density}

Referring to the expression of the full potential given by Eq.~(\ref{eq:potential}), we write the density in the same form,
\begin{equation}
\begin{aligned} 
    n(\bm r)   
    & = n^0(\bm r) + \sum_{R} \{ n^1_{R}(\bm r) - n^2_{R}(\bm r) \},
\end{aligned}
\label{Eq:chargeRho}
\end{equation}
where ${n}_0$ denotes a smooth density, tabulated on a real-space uniform grid over the whole unit cell. $n^{1}_{R}(\bm r)$ and  $n^2_{R}(\bm r)$ are true and smooth local terms defined only inside $R$ atomic sphere, which are expanded in spherical harmonics up to an angular momentum cutoff $l^{\rho}_{high}$, namely, $n^{1}_{R}(\bm r_R) =\sum_{L}n^{1}_{RL}(r_R)Y_{L}(\hat{\bm r}_R)$, $n^2_{R}(\bm r_R)=\sum_{L}n^2_{RL}(r_R)Y_{L}(\hat{\bm r}_R)$. Here, ${n}^2_{R}$ cancels out the ${n}^0$ within $s_R$.

Inside MT spheres, the true density is composed of both core and valence density,namely $n^1_{R}(\bm r) = n_{R}^{core}(\bm r) + n_{R}^{1,val}(\bm r)$. The core density $ n_{R}^{core}(\bm r)$ is usually calculated using atomic-like boundary conditions for the core states, 
\begin{equation} \label{core}
    n^{core}_{R}(\bm r) = \frac{1}{4\pi} \sum_{nl}^{core} (2l+1) |\psi_{Rnl}^{core}(r_R)|^2,
\end{equation}
 where $\psi_{Rnl}^{core}(r)$ is the radial amplitude of core wavefunction for the closed $nl$-shell normalized to unity inside the MT sphere. 
 At present, the core levels and core wavefunction are computed by solving the scalar-relativistic equation.   

For the the valence density, it can be obtained by,
\begin{equation}
\begin{aligned}
    n^{val}(\bm r)      & =  \sum_{ R L R' L'}  \chi_{ RL}(\bm r) \chi_{R'L'}^*(\bm r) D_{R L, R'L'}   \\
    & \equiv n^0(\bm r) + \sum_{R_1} \{ n_{R_1}^{1,val}(\bm r) - n_{R_1}^2(\bm r) \}
\end{aligned}
\label{Eq:ValenceRho}
\end{equation}
where $D$ denotes  density matrix, which can be usually given as, by solving the generalized eigen problem of $O$ and $H$ for the eigenvalue $\epsilon_i$ and eigenvector $\varphi_i$ ,
\begin{equation}
\begin{aligned} 
    D & = \sum_{i} f(\epsilon_i,\mu) \ket{\varphi_i}\bra{\varphi_i} 
\end{aligned}
\label{Eq:densitymtx1}
\end{equation}
or equivalently by caclualting the Green's function $G(z)=(zO-H)^{-1}$,
\begin{equation}
\begin{aligned} 
    D & = -\frac{1}{\pi} Im \int^{\infty} f(\epsilon,\mu) G(\epsilon^+) d\epsilon
\end{aligned}
\label{Eq:densitymtx2}
\end{equation}
where \( f(\epsilon_i, \mu) \) is the Fermi-Dirac function,  $\mu$ is the chemical potential. 
As a result, the first term of Eq.~(\ref{Eq:ValenceRho}) is written as, 
\begin{equation}
\begin{aligned} 
n^{0}(\bm r) & =   \sum_{RL,R'L'} \Psi^{\alpha}_{RL}(\bm r)\Psi^{*\alpha}_{R'L'}(\bm r) D_{RL,R'L'},
\end{aligned}
\end{equation} 
yielding a smooth $n^0$ on the uniform grid.
 The true valence density inside MT spheres,  namely $n_{R_1}^{1,val}(\bm r)$ in Eq.~(\ref{Eq:ValenceRho}), can be formulated as,
\begin{equation}
\begin{aligned}
 n_{R_1}^{1,val}(\bm r) = &   \sum_{RL,R'L',L_2L_3}[ \Phi^{\alpha,RL}_{R_1L_2} (\bm r) \Phi^{* \alpha,R'L'}_{R_1L_3}(\bm r)  ]  D_{R L, R'L'}    \\
& = \sum_{L_1} n^{1,val}_{R_1L_1}(r)Y_{L_1}(\bm{\hat r}_{R_1}),
\end{aligned}
\end{equation}
and
\begin{equation}
\begin{aligned}
n^{1,val}_{R_1L_1}(r) = \sum_{RL,R'L'} D_{R L, R'L'} 
[\sum_{L_2L_3} \Phi^{\alpha,RL}_{R_1L_2} \Phi^{* \alpha,R'L'}_{R_1L_3} C_{L_1 L_2 L_3}],\\
\end{aligned}
\end{equation}
where the functions $\Phi^{\alpha}$ are defined in Eq.~(\ref{Phi-Define}). Similarly, $n^2_{R_1}(\bm r)$ in Eq.~(\ref{Eq:ValenceRho}) can  be calculated in the same form inside MT spheres, simply by replacing $\Phi^{\alpha}$ with $\tilde{\Psi}^{\alpha}$ in Eq.~(\ref{smotthPSi-Define}).

\subsection{Effective Potential} \label{Sec-EffV}
The full potential $V(\bm r)$ in Eq.~(\ref{eq:potential}) comprises the electrostatic potential $V^{es}(\bm r)$ and the exchange-correlation potential $V^{xc}(\bm r)$, namely,
\begin{equation}
    V(\bm r) =V^{es}(\bm r) + V^{xc}(\bm r).
\end{equation}
In the following, we will present the computation of $V^{es}(\bm r)$ and $V^{xc}(\bm r)$ in the MT geometry.

\subsubsection{Electrostatic Potential}

The calculation of the electrostatic potential $V^{es}$ depends on the charge density throughout the unit cell, and thus involves the accurate treatment of the potential associated with fast changing true electron density  $n^1_{R}(\bm r)$  inside the muffin-tin spheres.  
Weinert introduced the pseudo-charge method~~\cite{weinert1981solution,methfessel2000full} to address this issue. It relies on the key insight that multiple charge densities $\rho$, that produce identical multipole moments inside MT spheres, can  generate the same electrostatic potential within the interstitial region. Hence,  we can define the local charge density $n_R$ inside a sphere $R$, which contains valence densiy, core density and nuclear charge $Z_R$, namely,
\begin{equation}
   n_R(\bm r_R) \equiv \{ n^1_{R}(\bm r_R)-n^2_{R}(\bm r_R)\} - Z_R \delta_{r_R=0}
\label{multipole1}
\end{equation}
which can generate the multipole moments,  
\begin{equation}
\begin{aligned}
q_{RL}[\rho] = \int_{r_R \le s_R} n_R(\bm r_R) r_R^l Y_{L} (\hat{\bm r}_R) d^3\bm r_R
\end{aligned}\label{multipole2}
\end{equation}
%
%
%

%
%
To accurately calcualte the electrostatic potentail inside interstitial region, one can introduce a pseudo-density $\tilde{n}^0(\bm r)$ which has the same multipole moments inside augmentation sphere $R$ as the density $n_{R}$.
Then the pseudo charge density on uniform grid and a smooth radial density are defined as
\begin{equation}
\begin{aligned}
\tilde{n}^{0}(\bm r) & = {n}^{0}(\bm r) + \sum_{RL} q_{RL}\mathscr{G}_{RL}(\bm r),\\
\tilde{n}^2_{R}(\bm r) & = {n}^2_{R}(\bm r) + \sum_{L} q_{RL}\mathscr{G}_{RL}(\bm r),
\end{aligned}\label{pseudo-n0-n1R}
\end{equation}
where $\mathscr{G}_{RL}$ is a Gaussian of moment unity with angular momentum $L$, localized inside the MT sphere $R$ with a negligible tail. Here, we use  
\begin{equation}
\label{gaussians}
\mathscr{G}_L(\bm r) =A_l(\frac{a^2}{\pi})^{3/2} (2a^2)^l e^{-a^2r^2}r^l Y_L(\hat{\bm r}),
\end{equation}
where $A_l$ is the coefficient that normalizes $\mathscr{G}_L(\bm r)$ to unity.
which has the Fourier transformation $\mathscr{G}_L$ is $\hat{\mathscr{G}_L}(\bm G) = A_l (-i\bm G)^l e^{\frac{-G^2}{4a^2}} Y_{L}(\hat{\bm G})$. 
Then we solve the following three Poisson equations 
\begin{equation}
\begin{aligned}
  \nabla^2 \tilde{V}^{0,es} &= -8\pi \tilde{n}^0,
\end{aligned}
\label{Possion-V0}
\end{equation}
\begin{equation}
\begin{aligned}
  \nabla^2 {V}_{R}^{1,es} &= -8\pi {n}^1_{R} ,
\end{aligned}
\label{Possion-V1R}
\end{equation}
\begin{equation}
\begin{aligned}
  \nabla^2 \tilde{V}_{R}^{2,es}  & = -8\pi \tilde{n}^2_{R},
\end{aligned}
\label{Possion-V2R}
\end{equation}
to give the three components representation of $V^{es}$ of the system as follows,
\begin{equation}\label{eq:potential-es} 
   V^{es}(\bm r)=\tilde{V}^{0,es}(\bm r)+\sum_{R} \{V^{1,es}_{R}(\bm r_{R})-\tilde{V}^{2,es}_{R}(\bm r_{R}) \}.
\end{equation}
Here, $\tilde{V}^{0,es}(\bm r)$ presents the true potential in interstitial region, and $\sum_{R} V^{1,es}_{R}(\bm r_{R})$ is the true electrostatic potential inside MT spheres. Here $\tilde{V}^{0,es}(\bm r)$  is tabulated on a real-space mesh, the remaining terms are all represented on a radial grid. $\tilde{V}^{0,es}(\bm r)$ and $\tilde{V}_{RL}^{2,es}(\bm r_{R}) $ can cancel each other inside $s_R$ sphere.

Then, the Poisson equation of Eq.~(\ref{Possion-V0}) can be solved by the standard technique fast Fourier transformation (FFT), namely
\begin{equation}
\begin{aligned}
  \tilde{V}^{0,es}(\bm G) = \frac{8\pi \tilde{n}^0(\bm G)}{|\bm G|^2}, \quad for \ \bm G \neq 0
\end{aligned} \label{V0G}
\end{equation}
so that $\tilde{V}^{0,es}(\bm r) = FFT [\tilde{V}^{0,es}(\bm G)]$. Once the interstitial potential $\tilde{V}^{0,es}$ has been calculated, the Possion Eq.~(\ref{Possion-V1R}) and Eq.~(\ref{Possion-V2R})  can be obtained by solving the Dirichlet boundary value  problem on the sphere. Since the pseudo-potential $\tilde{V}^{0,es}$ is equal to the true potential in the interstitial region, the radial potential at $s$-sphere boundary can be calculated by the following formula
\begin{equation}\label{V-on-s}
V^{es}(\bm s_R) = \sum_{L} V^{es}_{L}(s_R) Y_{L}(\hat{\bm r}_{R}) ,
\end{equation}
\begin{equation}
V^{es}_{L}(s_R)= 4\pi i^l \sum_{\bm G} V(\bm G)  e^{i\bm G\cdot \bm R } j_l (Gs_R)Y^{*}_L(\hat{\bm G})  .
\end{equation}
 
Then, to obtain the potentials $V_{R}^{1,es}$  and  $V_{R}^{2,es}$, we rewrite for each MT sphere, 
\begin{equation}
  V^{1,es}_{R}(\bm r_R) = \sum_{L} V_{RL}^{1,es}(r_R) Y_{L}(\hat{\bm r}_R),
\label{V-ins}  
\end{equation}
\begin{equation}
 \tilde{V}^{2,es}_{R}(\bm r_R) = \sum_{L} \tilde{V}_{RL}^{2,es}(r) Y_{L}(\hat{\bm r}_R)
\label{V-ins}  
\end{equation}
which have the same value at the sphere boundary, namely $V_{RL}^{1,es}(s_R) = \tilde{V}_{RL}^{2,es}(s_R)=V^{es}_{L}(s_R)$. As a result, we can solve for the radial $V_{RL}^{1,es}(r_R)$, \cite{weinert1981solution} 
\begin{equation}
\begin{aligned} 
& V_{RL}^{1,es}(r_R)  = \{V^{es}_L(s_R)  + \frac{2Z_R}{s_R} \sqrt{4\pi} \delta_{L0}\}(\frac{r_R}{s_R})^l - \frac{2Z_R}{r_R} \sqrt{4\pi} \delta_{L0}\\
& + \frac{8\pi}{2l+1} \{\int_0 ^{s_R}r^{'2}_R dr'_R n^{1}_{RL}(r'_R) \{ \frac{r^l_<}{r^{l+1}_>} - \frac{r^l_R}{s^{2l+1}}  r^{'l}  \} 
\end{aligned}
\label{Ves-1R}
\end{equation}
where $r_{<}(r_{>})$ is the smaller (larger) of $r_R$ and $r_{R}'$, and $Z_{ R}$ the atomic number of the element centered at $\bm R$,  $\frac{2Z_R}{r_R} \sqrt{4\pi} \delta_{L0}$ is contributed by nuclei. The calculation of $\tilde{V}_{R}^{2,es}$ can be obtained in the same way,
\begin{equation}
\begin{aligned} 
& \tilde{V}_{RL}^{2,es}(r_R)  = {V}^{es}_L(s_R)  (\frac{r_R}{s_R})^l \\
& + \frac{8\pi}{2l+1} \{\int_0 ^{s_R}r^{'2}_R dr'_R \tilde{n}^{2}_{RL}(r'_R) \{ \frac{r^l_<}{r^{l+1}_>} - \frac{r^l_R}{s^{2l+1}}  r^{'l}  \} 
\end{aligned}
\label{Ves-1R}
\end{equation}


\subsubsection{Exchange-Correlation Potential} \label{XC-Section}
The exchange-correlation potential is a function of electron density $n(\bm r)$, 
\begin{equation} \label{XCfunctional}
  V^{xc}(\bm r) = F^{xc}[n(\bm r)],
\end{equation}
which also can be written in three-component form,
\begin{equation}\label{eq:potential-xc} 
   V^{xc}(\bm r)={V}^{0,xc}(\bm r)+\sum_{R} \{V_{R}^{1,xc}(\bm r_{R})-{V}^{2,xc}_{R}(\bm r_{R}) \}.
\end{equation}
Here, $V_{R}^{1,xc}(\bm r_R)$  and $V_{R}^{2,xc}(\bm r_R)$ are expanded in shperical harmonics, namely, 
\begin{equation}\label{VXCLMexp}
\begin{aligned}
V_{R}^{1,xc}(\bm r_{R}) &= \sum_{L} V^{1,xc}_{RL}(r_R) Y_{L}(\hat {\bm r}_{R}), \\
V_{R}^{2,xc}(\bm r_{R}) &= \sum_{L} V^{2,xc}_{RL}(r_R) Y_{L}(\hat {\bm r}_{R}),
\end{aligned}
\end{equation} 
in which l-cutoff takes $l^{\rho}_{high}$.
${V}^{0,xc}(\bm r)$, $V_{R}^{1,xc}(\bm r_{R})$ and ${V}_{R}^{2,xc}(\bm r_{R})$ are calculated by the respective density ${n^0}$, $n^1_{R}$ and ${n}^2_{R}$ on the respective uniform grid and radial grid, namely, 
\begin{equation}
    \begin{aligned}
        {V}^{0,xc}(\bm r) &= F^{xc}[n^0(\bm r)], \\
        {V}^{1,xc}_{R}(\bm r_{R}) &= F^{xc}[n^1_{R}(\bm r_{R})], \\
        {V}^{2,xc}_{R}(\bm r_{R}) &= F^{xc}[n^2_{R}(\bm r_{R})].
    \end{aligned}\label{Vxc1}
\end{equation}
The calculation of $V_{RL}^{1,xc}$ can $V_{RL}^{2,xc}$ for different $L$ in Eq.~(\ref{VXCLMexp}) requires special care. 
To do so, one can generate a set of evenly distributed points $(\theta_i,\phi_i)$ as follows, 
     \begin{align*}
      \theta_i&=\arccos\left[1-\left(i-\tfrac{1}{2}\right)\delta z\right], \\
      \phi_i&=(i-1)\delta\phi,
     \end{align*}
where $i=1,...,N (N=l^{\rho,2}_{high})$, and $\delta z=2/N$ and $\delta\phi=\pi(1-\sqrt{5})$. Within a sphere, for a given radial $r$, we form N real-space points $\bm r_i$ on which we calculate the $V_{R}^{1,xc}(\bm r_i)$ by using the XC functional. Then, the $V_{RL}^{1,xc}(r)$ can be obtained by, ($V_{RL}^{2,xc}(r)$ is obtained in the same way.)
\begin{equation}
\left[ 
\begin{aligned}
   V_{RL_1}^{1,xc}(r) \\ 
   V_{RL_2}^{1,xc}(r) \\ 
   \vdots \quad\quad \\ 
   V_{RL_{\rho}}^{1,xc}(r) 
\end{aligned}
\right] = 
\left[ 
\begin{aligned}
   &Y_{L_1}( \hat{\bm r}_1) &  \cdots \quad  &Y_{L_{\rho}}( \hat{\bm r}_1) \\
   &Y_{L_1}( \hat{\bm r}_2) &  \cdots  \quad &Y_{L_{\rho}}( \hat{\bm r}_2)\\
   &\quad \vdots &\vdots\  \quad &\quad \vdots \\
   &Y_{L_1}( \hat{\bm r}_n) &  \cdots \quad  &Y_{L_{\rho}}( \hat{\bm r}_n)
\end{aligned}
\right] ^{-1}
\left[
\begin{aligned}
   V_{R}^{1,xc}(\bm r_1) \\
   V_{R}^{1,xc}(\bm r_2) \\
   \vdots \quad \quad \\
   V_{R}^{1,xc}(\bm r_n)
\end{aligned}
\right].
\end{equation}

\subsection{Total Energy}
Within the Kohn-Sham density-functional formalism, the total energy is separated into the kinetic energy $T^s$ of non-interacting electrons, the electrostatic interaction energy $E^{es}$ of the whole system, and the exchange-correlation energy $E^{xc}$, namely, 
\begin{eqnarray}
   E_{tot} = T^s + E^{es} + E^{xc}.
\end{eqnarray}

\subsubsection{Kinetic Energy}

The kinetic energy $T^s$ includes valence and core contributions, namely, $T^s=T^{val}+T^{core}$, which can be written as,
\begin{equation}
\begin{aligned}
  T^{val}&= \sum_i \epsilon_{i}^{val}-\int{n^{val}(\bm r) V(\bm r)} d^3\bm r, \\
  T^{core}&= \sum_i \epsilon_{i}^{core}-\int{n^{core}(\bm r) V(\bm r)} d^3\bm r,
\end{aligned}\label{Eq:Ks-val-core}
\end{equation}
 where $V(\bm r)$ denotes the effective potential, $\epsilon_{i}^{val}$ and $\epsilon_i^{core} $ are valence and core electron eigen energies.  In practical implementation, we can calculate the sum of valence eigenvalues  $\sum_{i}\epsilon_{i}^{val}$ either by using the exact diagonalization method or by applying contour integrals of the physical Green's function.\cite{turek1997electronic,Vitos2007} 

\subsubsection{Electrostatic Energy}
The total electrostatic energy $E^{es}$ is composed of the Hartree energy $E_{ee}^{es}$,  electron-nucleus interaction $E_{ez}^{es}$ and nucleus-nucleus interaction $E_{zz}^{es}$, namely, $E^{es}  = E_{ee}^{es} +  E_{ez}^{es} + E_{zz}^{es}$, which are given by,  
\begin{equation}
\begin{aligned}
  &E_{ee}^{es} = \frac{1}{2} \int {n(\bm r) V^{es}_{e}}(\bm r) d^3r, \\
  &E_{ez}^{es} = \frac{1}{2} \int {n(\bm r) V^{es}_{z}(\bm r)}  d^3r  +  \frac{1}{2}  \sum_{R}Z_R V^{es}_{e}(r_R)|_{r_R\rightarrow 0}, \\
  & E_{zz}^{es} =\frac{1}{2} \sum_{R}  Z_{R} [V^{es}_z(r_R) - \frac{Z_{R}}{ r_{R}}]|_{r_R\rightarrow 0},
\end{aligned}
\end{equation}
in which the electrostatic potential $ V^{es}_{e}$ is the Hartree potential and  $V^{es}_{z}$ denotes nuclei potential.

\subsubsection{Exchange–Correlation Energy}
The exchange correlation energy $E^{xc}$ can be expressed as
\begin{eqnarray} \label{Energy-xc1}
\begin{aligned}
   E^{xc} &= \int n(\bm r)\epsilon^{xc}[n(\bm r)]d^3r  
\end{aligned}
\end{eqnarray}
where $\epsilon^{xc}$ denotes the exchange-correlation energy density, which is made into the three-component form, in analogy to $V^{xc}$,
\begin{equation}\label{eq:xc-energy-density} 
   \epsilon^{xc}(\bm r)=\epsilon^{0,xc}(\bm r)+\sum_{R} \{\epsilon_{R}^{1,xc}(\bm r_{R})-\epsilon^{2,xc}_{R}(\bm r_{R}) \}.
\end{equation}
where $\epsilon^{0,xc}(\bm r)$ is represented on uniform grid, $\epsilon_{R}^{1/2,xc}(\bm r_{R}) = \sum_{L} \epsilon_{RL}^{1/2,xc}( r_{R})Y_{L}(\hat{\bm r}_R)$ are represented on radial grid.  The terms $\epsilon^{0,xc}(\bm r)$, $\epsilon_{R}^{1,xc}(\bm r_{R})$ and $\epsilon^{2,xc}_{R}(\bm r_{R})$ are calcualted with the respective ${n^0}$, $n^1_{R}$ and ${n}^2_{R}$ in the same way as $V^{xc}(\bm r)$ described in Sec.\ref{XC-Section}.  

\begin{figure}[htbp]
    \centering
    \includegraphics[scale=0.75]{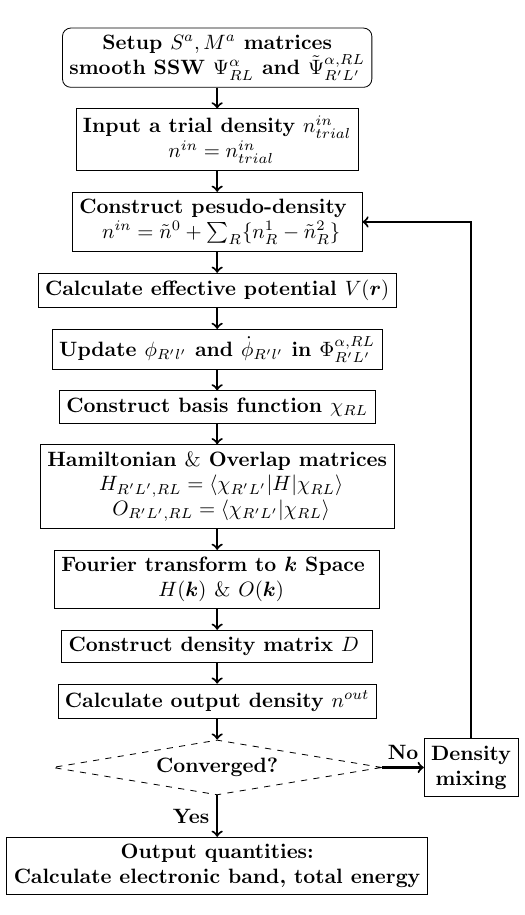} %
\caption{Flow chart of the FP-TB-LMTO based all-electron DFT self-consistent loop for electronic structure calculation.}
\label{Fig:Scf-loop}
\end{figure}

\begin{figure*}[htbp]
    \centering
    \includegraphics[scale=0.84]{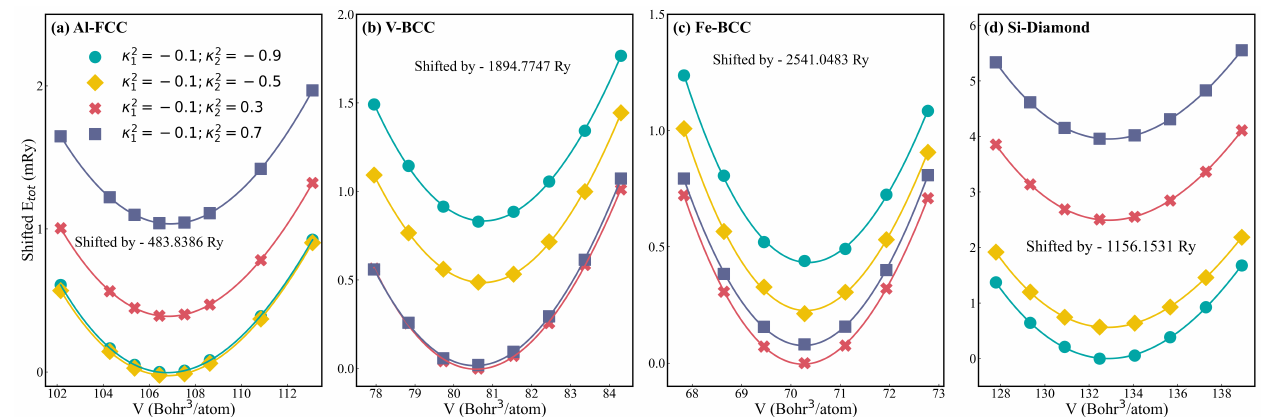} %
    \caption{Total energy (shifted by a constant) vs. volume (per atom) for (a) Al-FCC  (b) V-BCC (c) Fe-BCC and (d) Si-Diamond (with vacuum spheres) using different sets of $\kappa_1^2$ and $\kappa_2^2$. $s_R = 0.8 \omega$ is used for Al and Si, and $s_R = 0.85 \omega$ is used for V and Fe. 
} 
    \label{fig:SSW-robu}
\end{figure*}

\subsection{Self-Consistent Implementation}
In  this subsection, we describe the procedure for implementing the self-consistent all-electron FP-TB-LMTO method, as schematically shown in Fig.~\ref{Fig:Scf-loop}. The full self-consistency is achieved if both density $n(\bm r)$ and $E_{tot}$ are converged. For a given atomic lattice structure, the self-consistent calculation start with calculating the slope matrix $S^a$ and $M^a$, and then constructing the auxiliary augmented function $\Psi^{\alpha}_{RL}$ in Eq.~(\ref{eq:auxiliarySSW1}) on a uniform real-space grid, and the auxiliary augmentation function $\tilde{\Psi}_{R'L'}^{\alpha,RL}$ in Eq.~(\ref{eq:auxiliarySSW1}) on the radial grid inside MT sphere.It should be noted that $\Psi^{\alpha}_{RL}$ and $\tilde{\Psi}_{R'L'}^{\alpha,RL}$ remains unchanged during the self-consistent iterations, and the self-consistent loop only updates the partial-wave augmentation $\Phi^{\alpha,RL}_{R'L'}(\bm r)$ for the new TB-LMTO $\chi$ in each iteration. Next, to start the calculation,  we adopt the superposition of free atom charge densities as the trial density $n_{trial}^{in}$, which is a reasonable starting point for achieving a fully converged $n(\bm r)$. Once the initial $n_{trial}^{in}(\bm r)$ is constructed, the following self-consistent loop can be executed, for a self-consistent calculation of electronic structure, \\
1.Initializing electron density $n^{in}(\bm r)=n^{in}_{trial}(\bm r)$ in the three-component form in Eq.~(\ref{Eq:chargeRho}).  \\
2.Constructing the pseudo smooth density $\tilde{n}^0$ in the full space  and smooth radial density $\tilde{n}^2_{R}$ inside muffin-tin spheres  by Eqs.~(\ref{multipole1})-(\ref{pseudo-n0-n1R}), with the input $n^{in}(\bm r)$. \\
3.Calculating the three components of effective potential \( V(\bm{r}) \) , including the electrostatic potential in Eq.~(\ref{eq:potential-es}) and the exchange-correlation potential in Eq.~(\ref{eq:potential-xc}). \\
4.Calculating partial waves $\phi_{Rl}$ and its dertivates $\dot{\phi}_{Rl}$for a set of $\epsilon_{l}$ inside the muffin-tin sphere at $R$, by solving the scalar-relatitistic radial schrodinger equation.\\
5.Updating MTOs $\chi_{RL}$ in Eq.~(\ref{eq:ASSW}) by the new augmentation functions $ \Phi^{\alpha,RL}_{R'L'}$ with the above partial waves, while $\Psi^{\alpha}_{RL}$ and $\tilde{\Psi}_{R'L'}^{\alpha,RL}$ remain unchanged. \\
6.Computing Hamiltonian matrix $H_{R'L',RL}$ in Eqs.~(\ref{V-Matrix-elements}) and (\ref{T-Matrix-elements}), and overlap matrix $O_{R'L',RL}$ in Eq.~(\ref{O-Matrix-elements}).\\
7.Fourier transforming the Hamiltonian and overlap matrices to $\bm k$-space to obtain $H(\bm k)$ and $O(\bm k)$.\\
8.Calculating the density matrix $D$ by diagonalization or using Green's function method for each $\bm k$ and integrating over the  BZ.\\ 
9.Constructing the output density $n^{out}(\bm r)$ by calculating valence electron density with the density matrix $D$  in  Eq.~(\ref{Eq:ValenceRho}) and core electron density with Eq.~(\ref{core}). \\
10.Checking for convergence:  Evaluate the convergence criteria: \\
\textbullet \quad  checking if $max|\frac{n^{out}-n^{in}}{n^{in}}|\le n_{tol}$, where $n_{tol}$ is the density convergence threshold. \\
\textbullet  \quad checking if the total energy change $\Delta E_{tot} \le E_{tol}$, where $E_{tol}$ is the energy convergence threshold. \\
If both criteria are met, exit the self-consistent loop to output the total energy, electronic band structure and other physical quantities. Otherwise,we mix the density using the mixing procedure to generate a new $n(\bm r)$ and return to step 2 for another iteration.

For self-consistency, the Anderson mixing algorithm~\cite{anderson1965iterative} and the exact diagonalization method are employed in the present implementation. In the present implementation of FP-TB-LMTO method, we have implemented the multi-kappa basis sets with the flexibility of choosing $\kappa^2>0$ or $\kappa^2<0$, and incorporate the crystal symmetry to accelerate the computation in BZ integration. FFT uses FFTW (the Faster Fourier Transform in the West) package,\cite{fftw05} the matrix operations, including diagonalization and inversion, are using LAPACK (Linear Algebra PACKage) package,\cite{lapack99} Fermi-Dirac smearing is used to improve convergence and accuracy.\cite{PhysRev.137.A1441}

\section{RESULTS and DISCUSSIONS}\label{results}
To demonstrate the accuracy for the bulk materials, in the next section, we calculate the electronic band, total energy, phase ordering, and elastic constants (ECs) for a variety of bulk systems including different normal metals, compounds, and silicon, and compare with the caluclations with the well established FP-LMTO method implemented in Questaal~\cite{pashov2020questaal}. 
 We set $l_{high}=8$ for converging the augmentation in Eqs.~(\ref{eq:auxiliarySSW}) and (\ref{eq:ASSW}), and employ $l_{max}^{\rho}=4$ and $l_{max}^{V}=4$ for converging the expansion of potential and charge density inside the MT spheres in Eq.~(\ref{eq:potential}) and Eq.~(\ref{Eq:chargeRho}). Generally, the larger the potential sphere radius $s_R$ is used, the larger $l_{high}$ is required. In this work, for constructing the SSW, we use the screening hard sphere radius $a_R = 0.7 \omega$ for the calculated FCC, BCC, SC and Diamond (with vaccum spheres) structures, where $\omega$ denotes the average Wigner-Seitz-cell radius (WSA). For ensuring the high accuracy in total energy, we choose $R_{cutoff}= 3.8$~$\omega$ for the real-space numerical representation of $\Psi^{\alpha}_{RL}$ of Eq.~ (\ref{eq:auxiliarySSW}). All the calculations are performed by double-$\kappa^2$ TB-LMTO bases to ensuring the completeness in interstial region (in some case, the third $\kappa^2$ basis is used to treat the semicore electrons), and the partial wave inside MT sphere is solved on a single energy center $\epsilon_l$ for each l ($l\le l_{high}$) ($\epsilon_l$ is determined as the averaged band energy of l-component). For converging the BZ and real-space integration,
we use $16\times 16 \times 16$ $\bm k$-mesh for BZ integration and $16\times 16 \times 16$  uniform grid mesh for the BCC, FCC and SC structures,, we use  $12\times 12 \times 12$  $\bm k$-points and  $24\times 24\times 24$  real-space grid for Diamond structure, and $24\times 24 \times 20$  $\bm k$-points and  $24\times 24\times 32$ uniform grid mesh for HCP structures. For the calculation of ECs, we may increase the densities of the uniform grid and the $\bm k$ mesh to ensure the correct tiny energy response to the small structural deformation (e.g. $\le 0.1 meV$ for calculating the $C_{44}$ in V). 
All the calculations in this work are performed within the local spin density approximation (LSDA), with the von Barth and Hedin (vBH) parameterization for the exchange-correlation functional~\cite{vonbarth1972xc}. The FP-TB-LMTO and FP-LMTO methods use the same settings for comparison.  

\begin{figure*}[htbp]
    \centering
    \includegraphics[scale=0.900]{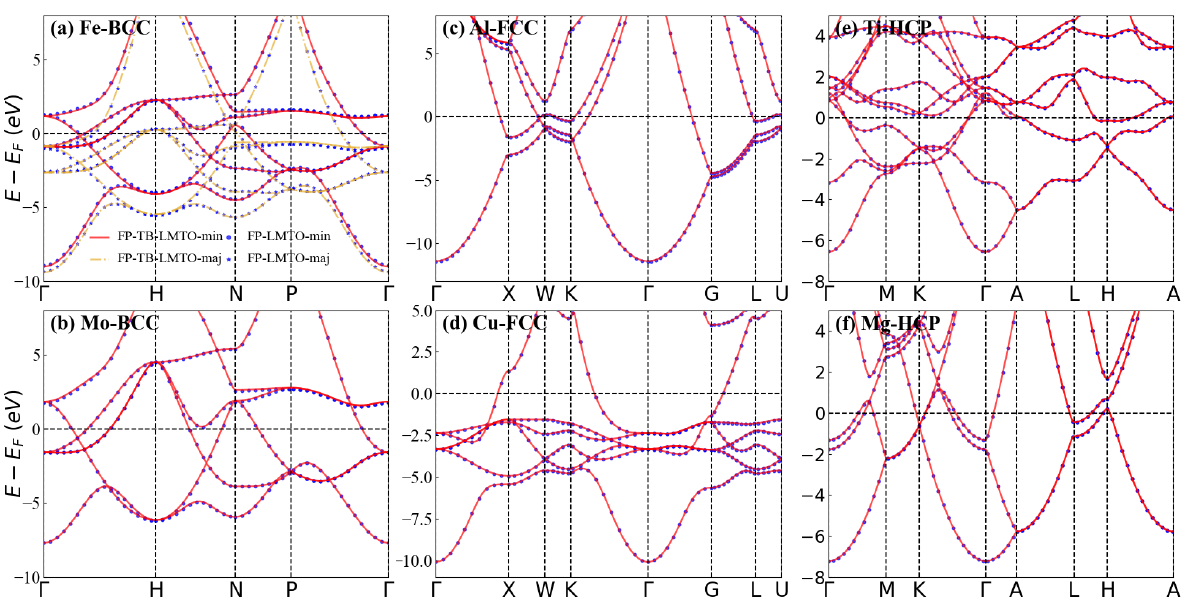} %
    \caption{Band structures for different metals with the specified $\omega$:
    (a)~Fe~BCC~($2.55$ Bohr), (b)~Mo~BCC~($2.86$ Bohr), (c)~Al~FCC~($2.94$ Bohr), (d)~Cu~FCC~($2.60$ Bohr), (e)~Ti~HCP~($2.88$ Bohr, $c/a = 1.60$), (f)~Mg~HCP~($3.26$ Bohr, $c/a=1.629$). The red solid and yellow dash-dot lines respectively represent the minority and majority spin results of FP-TB-LMTO, while the blue points and stars represent those from FP-LMTO (Questaal).  $s_R = 0.85 \omega$ is used for Fe, Mo,Ti and Mg; $s_R = 0.8 \omega$ is used for Al and Cu.
}
    \label{fig:Metal-Band}
\end{figure*}

\subsection{Robustness of Basis Functions}

In the present FP-TB-LMTO method, for the basis function $\chi_{RL} $ in Eq.~(\ref{eq:ASSW}) inside the MT sphere, the partial waves for $l\le l_{high}$ are self-consistently calculated at the band-energy centers, namely $\epsilon_l$, providing the optimal functions for expanding wavefunction inside MT spheres. The SSW in the interstitial region characterizes the TB-LMTO, remains unchanged in the self-consistent calculation, namely with the fixed $ \kappa^2$. The selection of $ \kappa^2$ for the SSWs determines the accuracy of the TB-LMTO basis in present calculations. The weak-dependence of the results on the selction of $\kappa^2$ can demonstrate the high robustness of TB-LMTO, presenting the easy usage of present method. In order to study the robustness of TB-LMTO method, Fig.~\ref{fig:SSW-robu} presents the the total energy vs.volume with different choice of $\kappa^2$ values for different systems including (a)~FCC~Al, (b)~BCC~V, (c)~BCC~Fe and (d)~Diamond~Si. Here, we investigate the dobule-$\kappa$ basis with the fixed $\kappa_1^2=-0.1$ Ry and different $\kappa^2_2=-0.9$, $-0.5$, $ 0.3$ and $0.7$ Ry, while the single-$\kappa$ basis is not considered due to its known low accuracy for the total energy calculation. 
As shown in Fig.~\ref{fig:SSW-robu}(a-d), the four bulk systems all exhibit almost the same equilibrium volume and response of total energy to volume changes for using different sets of basis, ensuring that all the chosen bases can produce the accurate physical properties. We can find, the energy difference between the results of $\kappa^2_2=-0.9$ Ry and $\kappa_2^2=0.7$ Ry, at the equilibrium volume, is 1.03~mRy in ~FCC~Al, 0.81~mRy in ~BCC~V, 0.36~mRy in BCC~Fe and 3.96~mRy in ~Diamond~Si. Therefore, we can see the difference in $\kappa_2^2$, as large as $1.6$ Ry, can only produce the total energy constant shift of few mRys in different systems, presenting the high robustness of double-$\kappa$ basis of TB-LMTO. It is found that the transition metals V and Fe with the postive vlaues of $\kappa_2^2=0.3$ and $0.7$ presents the lower total energy compared to the results of negative values of $\kappa_2^2$, while the simple metal Al and semiconductor Si with negative $\kappa_2^2$ shows the lower energies. This investigation suggests the optimal $\kappa_2^2$ should be positive values for transition metals due to the high kinetic energy of interstial electrons, and be negative for Al and semiconductor(or insulator) due to the relatively low kinetic energy of interstital electrons (which is determined the distance between $V_{MTZ}$ and valence band maximum).

\begin{figure*}[t]
    \centering
    \includegraphics[scale=0.900]{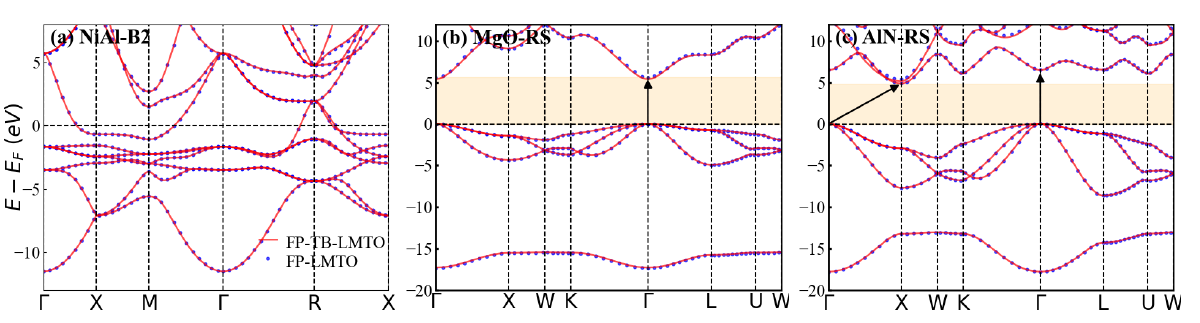} %
    \caption{Band structures for the different compounds with the specified WSA $\omega$:
    (a)~NiAl~B2~($2.64$ Bohr), (b)~MgO~RS~($2.42$ Bohr) and (c)~AlN~RS~($2.34$ Bohr). The red solid lines and bule points respectively represent the  results of FP-TB-LMTO and FP-LMTO (Questaal). $s_R=0.8 \omega$ is used for all spheres in the compound systems}
    \label{fig:Compound-Band}
\end{figure*}

\subsection{Electronic Band}
    To assess the accuracy and broad applicability of the FP-TB-LMTO method, this section presents calculations of the electronic bands for a variety of bulk materials, including normal metals, metallic and insulating compounds and semiconductor silicon. The electronic band structure is the basis for the interpretation of optical and electronic transport properties. 

 \subsubsection{Normal Metals}
 As shown in Fig.~\ref{fig:Metal-Band}, the electronic band structures along high-symmetry directions are presented for six bulk metals, including  BCC Fe~(a), BCC Mo~(b), FCC Al~(c), FCC Cu~(d), HCP Ti~(e) and HCP Mg~(f). The present results with the FP-TB-LMTO calculations are compared with the results by FP-LMTO (calculated with the well established Questaal electronic structure package). All the calculations are performed at specific system parameters $\omega$ and $s_R$ as shown in the caption of Fig.~\ref{fig:Metal-Band}.

As shown in Fig.~\ref{fig:Metal-Band}, the implemented FP-TB-LMTO (in solid lines) presents overall very good agreement with the calculations of FP-LMTO (in dots) in a wide range of energy for all the calculated bulk systems, demonstrating the correct implementation and important accuracy of the FP-TB-LMTO method. However, it is notable that there are small deviations between FP-TB-LMTO and FP-LMTO calculations in different $\bm k$ directions. For the HCP structure, the maximal devitation between the two calculations appears at the $\Gamma$ point below Fermi level with the absolute value of 0.011 eV and 0.007 eV in the respective Ti and Mg, which are negligible. Moreover, below the Fermi level, the maximum deviation reaches 0.032 eV in BCC Mo at the H point, are the values of 0.045 eV and 0.024 eV in the respective FCC Al and Cu at the $\Gamma$ point. In BCC~Fe with spin polarization, at the P point, FP-TB-LMTO presents the largest deviation with the value of 0.140~eV over FP-LMTO results,for the first band of majority spin below the Fermi level. These small diferences in band between the present FP-TB-LMTO and FP-LMTO can be majorly attributed to the difference in the envelope functions ultilized in the two MTOs: FP-TB-LMTO ultilizes the original Hankel function in the interstitial region, while in FP-LMTO, the smoothened Hankel function is constructed by convoluting with the Gaussian function.

%
%
%
%


\begin{figure}[t]
\includegraphics[scale=0.5]{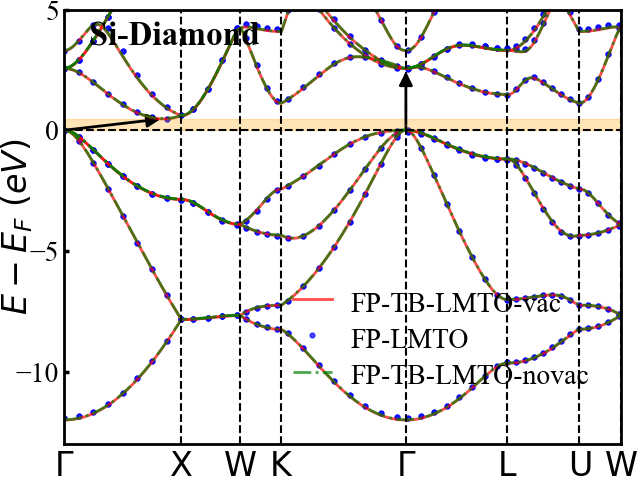}
\caption{\label{Fig:Si-band} The band structure of Si in diamond phase with lattice constant of 10.21 Bohr. The red solid line and green dash line represent the results with and without vacuum spheres from FP-TB-LMTO respectively, while the blue points represent the result without vacuum spheres from FP-LMTO (Questaal).}
\end{figure}
\subsubsection{Compounds and Silicon}

\begin{figure*}[t]
    \centering
    \hspace{0in}
    \includegraphics[scale=0.900]{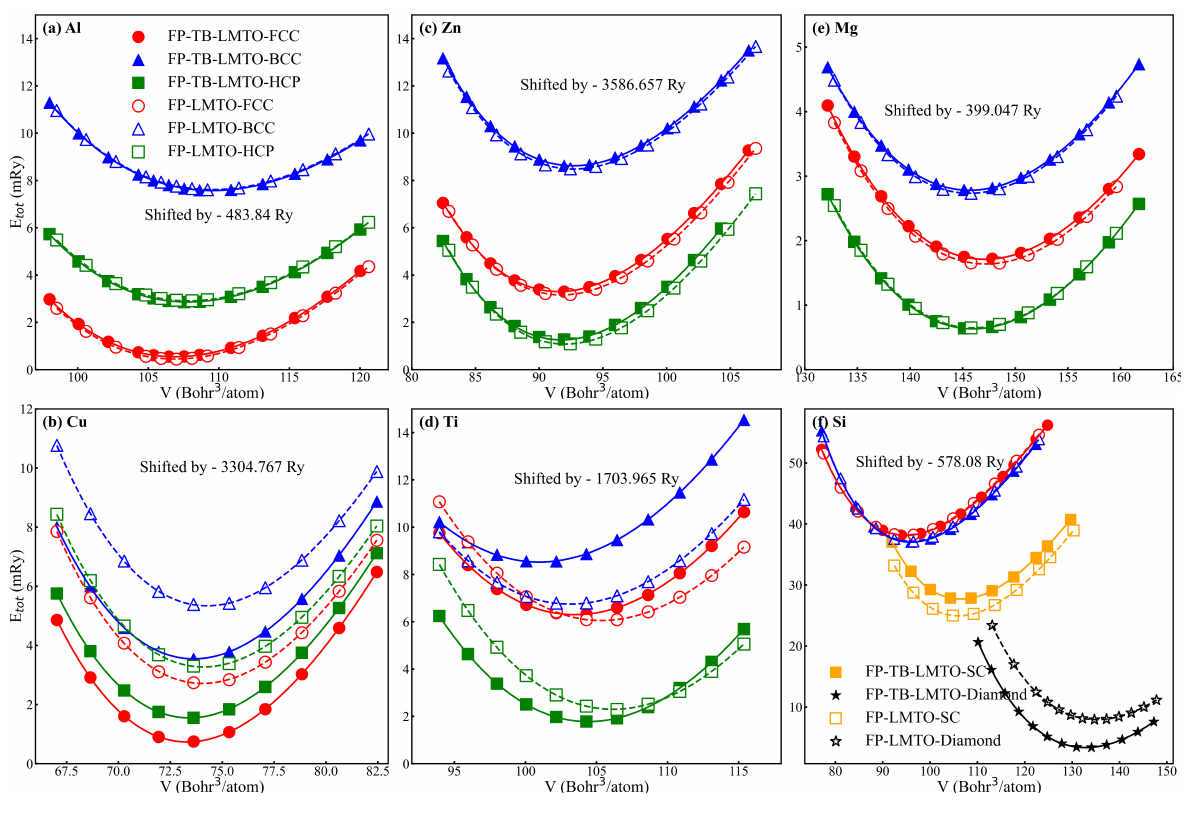  } %
    \caption{Total energy (shifted by a constant) vs.volume (per atom) for Al~(a), Cu~(b), Zn~(c), Ti~(d), Mg~(e) and Si~(f). Solid and open symbols correspond to FP-TB-LMTO and FP-LMTO (Questaal) results, respectively, while solid and dashed curves represent least-squares fits to the relation of BM3-EOS\cite{Birch1947}.  $s_R=0.8 \omega$ is used for Al,Cu,Zn and Si, and  $s_R=0.85 \omega$ is used for Ti and Mg.}
    \label{fig:Phase-Ordering}
\end{figure*}

Fig.\ref{fig:Compound-Band} presents the band structures for the compounds B2 NiAl~(a), Rock-salt (RS) MgO~(b) and RS AlN~(c) calculated by FP-TB-LMTO and FP-LMTO methods (the system parameters can be found in the caption). As shown, the present FP-TB-LMTO (solid line) calculations agree very well with the results of FP-LMTO  (blue dots) for all the compounds, demonstrating the accurate implementation of FP-TB-LMTO method. 
For the metallic NiAl in Fig.~\ref{fig:Compound-Band}(a), the maximum deviation in band energy between the two methods are below 10~meV, presenting excellent agreement (in the whole energy range as plotted) between FP-TB-LMTO and FP-LMTO calculations.   
For MgO in Fig.~\ref{fig:Compound-Band}(b), the deviation in valence band is quite small, while the deviation in the conduction band is noticeable, for example, at the $\Gamma$ point, the maximal energy deviation in conduction band bottom is 0.19~eV. As a result, the FP-TB-LMTO calcualtion presents a direct bandgap of  5.40 eV for MgO, compared to the result of FP-LMTO 5.59 eV. 
Along the $\Gamma$-X direction, the FP-TB-LMTO method gives the effective mass -2.08 for the hole at the valence band maximum and 0.41 for the electron at the conduction band minimum, agreeing well with the corresponding values -2.02 and 0.42 of the FP-LMTO results. 
Fig.~\ref{fig:Compound-Band}(c) presents the energy band for the AlN in RS structure with the two methods. The maximum deviation between FP-TB-LMTO and FP-LMTO band results appears at the $X$ point with a deviation value of 0.15~eV for the second band in the conduction band, while the valence bands agree very well in all the calculated directions. For AlN, both methods predict a indirect band gap between the $\Gamma$ and $X$ points, with the gap value of 4.87 with FP-TB-LMTO and 4.89eV with FP-LMTO. At the $\Gamma$ point, the direct band gap is  6.48 with FP-TB-LMTO and 6.50 with FP-LMTO. 
Along the $\Gamma$-X direction, with the FP-TB-LMTO method, the hole effective mass is -0.9 at the valence band maximum and the electron effective mass is 0.59 at the conduction band minimum, agreeing well with the corresponding values -0.93 and 0.60 of FP-LMTO results.

The band structure of semiconductor silicon in diamond structure is shown in Fig.~\ref{Fig:Si-band} for both FP-TB-LMTO (with and without vacuum spheres) and FP-LMTO calculations. It is clear that, in present implementation of FP-TB-LMTO, the resutls with (in red solid) and without (in green dotted line) show the excellent matching with each other. It is noticed that FP-TB-LMTO calculation presents a notable deviation at the bottom of valence band at the $\Gamma$ point, with a value of 0.09~eV lower than that of FP-LMTO (Other points with small deviation can be found in the conduction band.). As shown, both calculations illustrate an indirect band gap for diamond semiconductor, with the conduction band minimum lying close to $X$ in the direction $\Gamma-X$.  The FP-TB-LMTO method predicts the indirect band gap of 0.47 eV (in LDA) with good agreement with the value 0.48 eV of FP-LMTO calculation, demonstrating good consistency between the two methods. 
Along the $\Gamma$-X direction, the hole and electron effective masses at the respective valence band maximum and conduction band minimum are -0.32 and 0.94 by the FP-TB-LMTO, compared well to the corresponding values -0.33 and 0.91 with the FP-LMTO method.

\subsection{Phase Ordering}

\begin{table*}[t]
\caption{\label{tab:table3}Equilibrium lattice constant $a$, elastic constants $C_{ij}$, bulk moduli $B$, and magnetic moment per atom $m$ are calculated using FP-TB-LMTO (in bold), compared with FP-LMTO (in parentheses). The unit of $a$ is in Bohr, the units of $C_{ij}$ and $B$ are in GPa, and the unit of $m$ is in Bohr magneton $\rm \mu_B$. }
\begin{ruledtabular}
\begin{tabular}{ccccccccccc}
 \ & $a$ &$c/a$ &$C_{11}$&$C_{12}$&$C_{44}$&$C_{13}$&$C_{33}$ 
 &$B$ & $m$ \\ \hline
 Be(HCP)&$\textbf{4.21}$\scriptsize{($4.21$)}
 &$\textbf{1.579}$\scriptsize{($1.579$)} &$\textbf{336}$\scriptsize{($\textsl{336}$)}&$\textbf{17}$\scriptsize{($\textsl{27}$)}&$\textbf{180}$\scriptsize{($\textsl{172}$)}&$\textbf{15}$\scriptsize{($\textsl{15}$)}&$\textbf{401}$\scriptsize{($\textsl{400}$)}
 &$\textbf{131}$\scriptsize{($\textsl{131}$)} &  \\ 
 %
 %
 Mg(HCP)&$\textbf{5.92}$\scriptsize{($5.92$)}
 &$\textbf{1.629}$\scriptsize{($1.629$)} &$\textbf{86}$\scriptsize{($\textsl{74}$)}&$\textbf{18}$\scriptsize{($\textsl{24}$)}&$\textbf{24}$\scriptsize{($\textsl{19}$)}&$\textbf{18}$\scriptsize{($\textsl{22}$)}&$\textbf{80}$\scriptsize{($\textsl{73}$)}
 &$\textbf{39}$\scriptsize{($\textsl{40}$)} &  \\ 
 Si(novac) &$\textbf{10.26}$\scriptsize{($\textsl{10.26}$)}&  &$\textbf{143}$\scriptsize{($\textsl{157}$)}&$\textbf{64}$\scriptsize{($\textsl{62}$)}&$\textbf{87}$\scriptsize{($\textsl{100}$)}& &  
 &$\textbf{90}$\scriptsize{($\textsl{93}$)} &  \\  
  Si(vac) &$\textbf{10.21}$ &  &$\textbf{166}$ &$\textbf{59}$ &$\textbf{112}$ & &  
 &$\textbf{95}$  &  \\  
 %
 %
  %
 %
 Al(FCC)&$\textbf{7.53}$\scriptsize{($\textsl{7.53}$)}&  
 &$\textbf{123}$\scriptsize{($\textsl{121}$)}&$\textbf{62}$\scriptsize{($\textsl{64}$)}&$\textbf{37}$\scriptsize{($\textsl{38}$)}& &  
 &$\textbf{82}$\scriptsize{($\textsl{83}$)} & \\   
 Ti(HCP)&$\textbf{5.34}$\scriptsize{($5.36$)}
 &$\textbf{1.588}$\scriptsize{($1.588$)} &$\textbf{193}$\scriptsize{($\textsl{187}$)}&$\textbf{62}$\scriptsize{($\textsl{67}$)}&$\textbf{29}$\scriptsize{($\textsl{31}$)}&$\textbf{76}$\scriptsize{($\textsl{77}$)}&$\textbf{193}$\scriptsize{($\textsl{189}$)} 
 &$\textbf{112}$\scriptsize{($\textsl{112}$)} &  \\ 
 V(BCC)&$\textbf{5.45}$\scriptsize{($5.48$)}&  
&$\textbf{307}$\scriptsize{($\textsl{306}$)}&$\textbf{121}$\scriptsize{($\textsl{123}$)}&$\textbf{17}$\scriptsize{($\textsl{16}$)}& & 
 &$\textbf{183}$\scriptsize{($\textsl{184}$)} &  \\  
 Fe(BCC)&$\textbf{5.18}$\scriptsize{($5.18$)}&  
 &$\textbf{339}$\scriptsize{($\textsl{371}$)}&$\textbf{173}$\scriptsize{($\textsl{187}$)}&$\textbf{148}$\scriptsize{($\textsl{165}$)}& & 
 &$\textbf{232}$\scriptsize{($\textsl{250}$)} & $\textbf{2.04}$\scriptsize{($\textsl{1.98}$)} \\  
 Co(HCP)&$\textbf{4.57}$\scriptsize{($4.60$)}&  $\textbf{1.620}$\scriptsize{($1.610$)} 
 &$\textbf{495}$\scriptsize{($\textsl{467}$)}&$\textbf{175}$\scriptsize{($\textsl{177}$)}&$\textbf{128}$\scriptsize{($\textsl{124}$)}&$\textbf{127}$\scriptsize{($\textsl{135}$)}&$\textbf{538}$\scriptsize{($\textsl{512}$)}
 &$\textbf{265}$\scriptsize{($\textsl{260}$)} & $\textbf{1.53}$\scriptsize{($\textsl{1.52}$)} \\ 
 Ni(FCC)&$\textbf{6.47}$\scriptsize{($6.47$)}&  
 &$\textbf{351}$\scriptsize{($\textsl{354}$)}&$\textbf{199}$\scriptsize{($\textsl{196}$)}&$\textbf{183}$\scriptsize{($\textsl{169}$)}& & 
 &$\textbf{246}$\scriptsize{($\textsl{248}$)} & $\textbf{0.60}$\scriptsize{($\textsl{0.60}$)} \\  
 %
 %
  Cu(FCC) &$\textbf{6.67}$\scriptsize{($\textsl{6.67}$)}&  &$\textbf{230}$\scriptsize{($\textsl{228}$)}&$\textbf{167}$\scriptsize{($\textsl{165}$)}&$\textbf{108}$\scriptsize{($\textsl{101}$)}& & 
 &$\textbf{188}$\scriptsize{($\textsl{186}$)} &  \\    
 Zn(HCP)&$\textbf{4.86}$\scriptsize{($4.87$)}& $\textbf{1.845}$\scriptsize{($1.843$)}
 &$\textbf{271}$\scriptsize{($\textsl{267}$)}&$\textbf{109}$\scriptsize{($\textsl{109}$)}&$\textbf{51}$\scriptsize{($\textsl{48}$)}&$\textbf{98}$\scriptsize{($\textsl{104}$)}&$\textbf{105}$\scriptsize{($\textsl{104}$)}
 &$\textbf{105}$\scriptsize{($\textsl{104}$)} &   \\ 
Mo(BCC)&$\textbf{5.80}$\scriptsize{(5.83)} & &$\textbf{519}$\scriptsize{($\textsl{503}$)}&$\textbf{115}$\scriptsize{($\textsl{122}$)}&$\textbf{120}$\scriptsize{($\textsl{99}$)}& 
 &   &$\textbf{250}$\scriptsize{($\textsl{249}$)} &    \\ 
Tc(HCP)&$\textbf{5.08}$\scriptsize{($5.10$)} &$\textbf{1.597}$\scriptsize{($\textsl{1.593}$)}&$\textbf{595}$\scriptsize{($\textsl{551}$)}&$\textbf{179}$\scriptsize{($\textsl{197}$)}&$\textbf{180}$\scriptsize{($\textsl{161}$)}&$\textbf{158}$\scriptsize{($\textsl{167}$)}
 &$\textbf{620}$\scriptsize{($\textsl{584}$)}  &$\textbf{310}$\scriptsize{($\textsl{305}$)} &    \\ 
 Rh(FCC)&$\textbf{7.03}$\scriptsize{($\textsl{7.08}$)}&  
 &$\textbf{505}$\scriptsize{($\textsl{484}$)}&$\textbf{202}$\scriptsize{($\textsl{208}$)}&$\textbf{264}$\scriptsize{($\textsl{247}$)}& &  
 &$\textbf{303}$\scriptsize{($\textsl{300}$)} &  \\   
 Pd(FCC)&$\textbf{7.23}$\scriptsize{($\textsl{7.27}$)}&  
 &$\textbf{268}$\scriptsize{($\textsl{258}$)}&$\textbf{201}$\scriptsize{($\textsl{197}$)}&$\textbf{98}$\scriptsize{($\textsl{86}$)}& &  
 &$\textbf{223}$\scriptsize{($\textsl{217}$)} &  \\   
 Ag(FCC)&$\textbf{7.55}$\scriptsize{($\textsl{7.58}$)}  &  &$\textbf{169}$\scriptsize{($\textsl{164}$)}&$\textbf{123}$\scriptsize{($\textsl{123}$)}&$\textbf{70}$\scriptsize{($\textsl{64}$)}& &  
 &$\textbf{138}$\scriptsize{($\textsl{137}$)} &  \\  
Cd(HCP)&$\textbf{5.53}$\scriptsize{($5.54$)} &$\textbf{1.849}$\scriptsize{($\textsl{1.840}$)}&$\textbf{203}$\scriptsize{($\textsl{158}$)}&$\textbf{114}$\scriptsize{($\textsl{72}$)}&$\textbf{26}$\scriptsize{($\textsl{26}$)}&$\textbf{78}$\scriptsize{($\textsl{64}$)}
 &$\textbf{73}$\scriptsize{($\textsl{77}$)}  &$\textbf{73}$\scriptsize{($\textsl{74}$)} &    \\ 
 Pt(FCC)&$\textbf{7.34}$\scriptsize{($\textsl{7.38}$)}  &  &$\textbf{346}$\scriptsize{($\textsl{324}$)}&$\textbf{273}$\scriptsize{($\textsl{276}$)}&$\textbf{75}$\scriptsize{($\textsl{81}$)}& &  
 &$\textbf{297}$\scriptsize{($\textsl{292}$)} &  \\  
 %
 Au(FCC) &$\textbf{7.65}$\scriptsize{($\textsl{7.68}$)}  &  &$\textbf{219}$\scriptsize{($\textsl{217}$)}&$\textbf{176}$\scriptsize{($\textsl{173}$)}&$\textbf{68}$\scriptsize{($\textsl{65}$)}& &  
 &$\textbf{190}$\scriptsize{($\textsl{188}$)} &  \\    
 NiAl(B2) &$\textbf{5.38}$\scriptsize{($\textsl{5.35}$)} &  &$\textbf{244}$\scriptsize{($\textsl{243}$)}&$\textbf{153}$\scriptsize{($\textsl{153}$)}&$\textbf{46}$\scriptsize{($\textsl{45}$)}& & 
 &$\textbf{183}$\scriptsize{($\textsl{183}$)} &  \\  
  MgO(RS) &$\textbf{7.82}$\scriptsize{($\textsl{7.83}$)} &  &$\textbf{355}$\scriptsize{($\textsl{359}$)}&$\textbf{78}$\scriptsize{($\textsl{77}$)}&$\textbf{149}$\scriptsize{($\textsl{158}$)}& & 
 &$\textbf{170}$\scriptsize{($\textsl{171}$)} &  \\ 
  AlN(RS) &$\textbf{7.58}$\scriptsize{($\textsl{7.58}$)} &  &$\textbf{514}$\scriptsize{($\textsl{480}$)}&$\textbf{151}$\scriptsize{($\textsl{169}$)}&$\textbf{366}$\scriptsize{($\textsl{336}$)}& & 
 &$\textbf{272}$\scriptsize{($\textsl{273}$)} &  \\  
\end{tabular}
\end{ruledtabular}
\end{table*}

Fig.\ref{fig:Phase-Ordering} shows the total energy versus volume per atom for BCC, FCC, and HCP phases of Al~(a), Cu~(b), Zn~(c), Ti~(d), and Mg~(e), and for BCC, FCC, simple cubic~(SC) and diamond phases of Si~(f), using the FP-TB-LMTO (in the filled) and FP-LMTO (in the empty) methods. For both methods, the total energy is shifted by the same constant as denoted in Fig.~\ref{fig:Phase-Ordering}. It is clear that, for all the calculated phases of Al, Zn and Mg, the absolute energy of the present FP-TB-LMTO  matches very well to the FP-LMTO results, while some small deviations in the total energy between  two methods are exhibited in the different phases of Cu,Ti and Si. We can evaluate the equilibrium total energy, volume, and bulk modulus by the third-order Birch-Murnaghan equation of state (BM3-EOS)~\cite{Birch1947}, namely
\begin{equation} \label{Birch-Murnaghan}\small
\begin{aligned}
& E(\mathcal{V}) = E_0 + \frac{9}{16}B_0 \mathcal{V}_0 \\
&\times \left\{ \left[\left(\frac{\mathcal{V}_0}{\mathcal{V}}\right)^{\frac{2}{3}} - 1 \right]^{3}B_0' - \left[\left(\frac{\mathcal{V}_0}{\mathcal{V}}\right)^{\frac{2}{3}} - 1\right]^{2} \left[4\left(\frac{\mathcal{V}_0}{\mathcal{V}}\right)^{\frac{2}{3}} - 6\right]\right\}
\end{aligned}
\end{equation}
where $E_0$, $B_0$, $\mathcal{V}_0$, and $B'_0$ are the total energy, bulk modulus, volume per unit cell at the zero pressure, and the derivative of bulk modulus with respect to the pressure, respectively. We here fit the results of FP-TB-LMTO (in solid line) and FP-LMTO (in dashed line) to the BM3-EOS.  It is clear that all the results of FP-TB-LMTO calculations can be smoothly fitted, presenting the important numerical stability of our present implmentation with the double augmentation scheme. For all the materials Al, Zn, Mg, Cu, Ti and Si, both FP-TB-LMTO and FP-LMTO methods correctly predict the equilibrium structures and phase ordering. In particular, for ground-state structure, Al and Cu adopt the FCC, Zn, Ti, and Mg stabilize in the HCP, and Si prefers the Diamond, all agreeing with the experimental observations. The FP-TB-LMTO calculations produce the equilibrium volume $\mathcal{V}_0$ and $B_0$ in high consistency with the results of FP-LMTO, for example, the FP-TB-LMTO presents the values $\mathcal{V}_0=106.9$, $73.3$, $91.9$, $104.4$, $146.0$ and $132.9$   Bohr$^3$/atom
(with the corresponding $B_0=82$, $188$, $105$, $112$,$40$ and $95$GPa) 
for the respective FCC Al, FCC Cu, HCP Zn, HCP Ti, HCP Mg and Diamond Si, compared to the corresponding results of FP-LMTO,
namely $\mathcal{V}_0=106.9$, $74.1$, $91.9$, $106.2$, $146.0$ and $135.2$~Bohr$^3$/atom 
(with the corresponding $B_0=83$, $186$, $104$, $112$, $40$ and $93$~GPa).
Moreover, for the HCP phase, FP-TB-LMTO and FP-LMTO predicts almost identical equilibrium $c/a$ values of  1.65,  1.65,  1.85,  1.588,  and 1.63 for  Al, Cu, Zn, Ti and Mg. 
In addition, it should be mentioned that the present FP-TB-LMTO predicts the lower equilibrium energies for FCC Cu, HCP Ti and Diamond Si than the FP-LMTO calculations, reflecting the differences between the implementations of FP-TB-LMTO and FP-LMTO. In particular, for the equilibrium energies, the FP-TB-LMTO produces the 1.98 mRy/atom in FCC Cu, 0.52 mRy/atom in HCP Ti, 4.58 mRy/atom in Diamond Si (in absolute total energy) lower than FP-LMTO results. The small deviations in absolute energy and structures between FP-TB-LMTO and FP-LMTO are attributed to different basis functions in the interstitial region, which is modified by the convolution with Gaussian in FP-LMTO implemented in Questaal.

\subsection{Elastic Constants}
Elastic constants (EC) characterize the ability of a material to deform under any small stresses. By employing the Voigt notation\cite{belytschko2014nonlinear}, the fourth-rank tensor  $C$ can be be  arranged in a $6 \times 6$ matrix with maximum 21 different elements. For small strain $\bm \eta = [e_1,e_2,e_3,e_4,e_5,e_6]^{T}$, Hooke’s law is valid and the crystal energy  $E(\mathcal{V},\bm \eta)$, can be expanded as a Taylor series,
\begin{eqnarray}
\begin{aligned}
   E(\mathcal{V},\bm \eta)&= E(\mathcal{V}_0,0) + \frac{\mathcal{V}_0}{2}\sum_{i,j=1}^6C_{ij}e_ie_j+ O({e_i^3}) 
\end{aligned} \label{Taylor}
\end{eqnarray}
where $C_{ij}$ is the element of EC, and $E(\mathcal{V}_0,0)$ is the energy of the unstrained system with equilibrium volume $\mathcal{V}_0$.

To further validate the accuracy of the FP-TB-LMTO method, we investigated the ECs for a wide range of materials, as shown in Table \ref{tab:table3}. ECs reflect the energy response of a material to small strains around its equilibrium volume. Their calculation requires highly precise method and thus provides a critical benchmark for testing the numerical implementation of the FP-TB-LMTO approach. According to symmetry considerations, three independent ECs (\(C_{11}\), \(C_{12}\) and \(C_{44}\)) are calculated for cubic crystals, while five constants (\(C_{11}\), \(C_{12}\), \(C_{13}\), \(C_{33}\) and \(C_{44}\)) are required for HCP crystals~\cite{WALLACE1970301}.  In the following calculation, volume-conserving deformations were applied in the primitive cell, and all the ECs are calculated by applying the deformation ranging from -1.5\% to 1.5\%.

The EC results in Table~\ref{tab:table3} show very good agreement between FP-TB-LMTO and FP-LMTO for various materials and crystal structures (with relative deviations generally within a few percent).For example,
in HCP Ti, the FP-TB-LMTO produces 193, 62, 76, 193 and 29 GPa for the respective $C_{11}$, $C_{12}$, $C_{13}$, $C_{33}$ and $C_{44}$ close to the corresponding values 187, 67 and 77, 189 and 31 GPa with FP-LMTO, where the relative deviation is small and acceptable for different methods. Moreover, for the compound AlN in rock-salt structure, the values of $C_{11}$,$C_{12}$ and $C_{44}$ are 514, 151 and 366~GPa with FP-TB-LMTO, in good agreement with the corresponding FP-LMTO results 480, 169, 336~GPa.  In addition, for the diamond Si, it is worth to mention that the FP-TB-LMTO calculation without vacuum sphere presents the noticeable underestimation for the $C_{11}$ and $C_{44}$, compared to the calculations with vacuum spheres. It is noticed that EC values of FP-LMTO calculations lies between the FP-TB-LMTO calculation with and without vacuum spheres, with acceptable deviation. For example, $C_{11}$=157 GPa with the FP-LMTO, while FP-TB-LMTO generates the value 143 GPa with vacuum spheres, and 166 GPa with vacuum spheres. It is worth to mention that, for V in BCC, the computation of $C_{44}$ is very chanllenging due to its very small magnitude. It is clear that the FP-TB-LMTO produces the $C_{44}=17$, compared very well with the result of FP-LMTO 16 Gpa. For $C_{44}$ of V, FP-TB-LMTO calculation (with semicore electron treated as valence) can correctly produce the $0.1$meV energy change under a deformation of 1.5\% out of the total energy about -1894.775 Ry.
Besides the ECs, Table~\ref{tab:table3} also include the equilibrium lattice constants, $a/c$ ratio for different HCP materials, and bulk modulus and magnetic momment (for Fe, Co, Ni). It is clear that all these strctural properties and magnetic moment calculated by the present FP-TB-LMTO are in high consistency with the FP-LMTO with the well developed Questaal package.

\section{CONCLUSIONS}\label{conclusion}
In summary, we have implemented the all-electron FP-TB-LMTO-based first-principles DFT method for self-consistent calculatiion of the electronic structure and total energy of materials. 
By introducing the double augmentation scheme, the SSW based MTO is accurately represented on a double grids, and then the accurate representation and calculations of full charge density, full potential, TB Hamiltonian and total energy are all achieved to realize the all-electron DFT simulation approach. 
We have domenstrated the high accuracy and robustness of FP-TB-LMTO method by calculating a wide variety of materials, including the total energy, band structure, phase ordering, equilibrium lattice constant and elastic constants,  in very good agreement with other well established FP method. With the important features of the complete,minimal,short-ranged and physically transparent basis, the present FP-TB-LMTO method for DFT provides an accurate approach for first-principles tight-binding simulation of solid-state materials and devices.
The presented algorithms and numerical implementations are starightforward to extend to realize the FP implementation of EMTO and NMTO method.

\begin{acknowledgments}
Y.K.acknowledges financial support from Foundation of National Key Laboratory of Computational Physics (grant.no.HX02021-22)  and NSFC (grant No.12227901). The authors thank the HPC platform of ShanghaiTech University for providing the computational facility.
\end{acknowledgments}


\appendix

\section{Connecting slope matrix $S^a$ and $M^a$} \label{App.StoM}
By defining Bessel functions $j_l(\kappa^2, r) \equiv \kappa^{-l} J_l(\kappa r)$ and  Neumann functions $n_l(\kappa^2, r) \equiv \kappa^{l+1} N_l(\kappa r)$ (where $J_l(\kappa r)$ and $N_l(\kappa r)$ denotes spherical Bessel and Neumann functions, respectively.),  the Hankel function $H_{L}(\kappa^2, \bm r)$ in Eq.~(\ref{eq:M-matrix}) can be expressed as,
    \begin{equation}
    \label{hankelfunc}
    \begin{aligned}    
         H_{L}(\kappa^2, \bm r) & \equiv -i \kappa^{l+1}  h_l^{(1)}(\kappa r)Y_{L}(\bm r) \\
         & = \kappa^{l+1}[N_l(\kappa r) - i J_l(\kappa r)] \\
         & \equiv  n_l(\kappa^2, r) - i \kappa^{2l+1}j_l(\kappa^2, r),
    \end{aligned}
    \end{equation}  
where $h^{(1)}(\kappa r) $ is the spherical Hankel function of the first kind. We apply the projection operator $ \mathscr{\hat{P}}$ to the   Eq.~(\ref{hankelfunc}), then we obtain,
    \begin{equation}
    \begin{aligned}
         \mathscr{\hat{P}}_{R'L'}(r) H_{L}(\kappa^2, \bm r_{R})  &=  n_{l'}(\kappa^2, r) \delta_{RR'}\delta_{LL'} \\
         & + j_{l'}(\kappa^2, r) B_{R'L',RL}(\kappa^2)     ,   
    \end{aligned}\label{ProjHankel}
    \end{equation} 
    where $B(\kappa^2)$ is the bare structure matrix~\cite{eyert2007envelope}. 

 Next, we proceed with expanding SSWs of Eq.~(\ref{eq:M-matrix}) into the Hankel functions. It is, however, simpler to derive the inverse expansion,
    \begin{equation}
    \label{M-inversion}
    \begin{aligned}
        \sum_{R'L'} \Psi^{\alpha}_{R'L'}(\kappa^2,\bm r) [M^a(\kappa^2)]_{R'L',RL}^{-1} = H_{L}(\kappa^2, \bm r_{R}).
    \end{aligned}
    \end{equation}
By applying the projection operator $\mathscr{\hat{P}}_{R''L''}(a_{R''})$ to Eq.~(\ref{M-inversion}) at the spheres' boundary, we have
    \begin{equation}
    \begin{aligned}
       &  \sum_{R'L'} \mathscr{\hat{P}}_{R''L''}(a_{R''}) \Psi^{\alpha}_{R'L'}(\kappa^2,\bm r) [M^a(\kappa^2)]_{R'L',RL}^{-1} \\
       & \quad = \mathscr{\hat{P}}_{R''L''}(a_{R''}) H_{L}(\kappa^2, \bm r_{R}).
    \end{aligned}
    \end{equation}
By combining the Eqs.~(\ref{ProjPhi}, \ref{ProjHankel}) and the boundary condition stated in Eq.~(\ref{boundary cond.}), we can  obtain,
    \begin{equation}
    \begin{aligned}
    \label{M-matrix1}
         [M^a(\kappa^2)]_{R''L'',RL}^{-1}  & = n_{l''}(\kappa^2, a_{R''}) \delta_{RR''}\delta_{LL''} \\
         & + j_{l''}(\kappa^2, a_{R''}) B_{R''L'',RL}(\kappa^2),
    \end{aligned}
    \end{equation}
then  we can find the relation in matrix notation,
    \begin{equation}
    \label{M-matrix2}
    \begin{aligned}
        M^a(\kappa^2) & = [n(\kappa^2,a) + j(\kappa^2,a)B(\epsilon)]^{-1} \\
        & = [\frac{n(\kappa^2,a)}{j(\kappa^2,a)} + B(\epsilon)]^{-1} \frac{1}{j(\kappa^2,a)}
    \end{aligned}
    \end{equation}
Next, by combining the Eqs.~(\ref{ProjPhi}) and (\ref{ProjHankel}), we can again apply the projector $\mathscr{\hat{P}}_{R''L''}(r_{R''})$ to Eq.~(\ref{M-inversion}) to obtain the relation, in matrix notation,
    \begin{equation}
    \begin{aligned}
        & [f^a(\kappa^2, r) + g^a(\kappa^2, r) S^a(\kappa^2)] M^{a}(\kappa^2)^{-1} \\
        & = n(\kappa^2, r) +  j(\kappa^2, r) B(\kappa^2).
    \end{aligned}\label{fgnjB}
    \end{equation}
With Eqs.~(\ref{boundary cond.}) and (\ref{fgjn}), we can find 
    \begin{equation}
          S^a(\kappa^2) M^a(\kappa^2)^{-1}  = \frac{1}{aj(\kappa^2, a)} + \frac{aj'(\kappa^2, a)}{j(\kappa^2, a)} M^a(\kappa^2)^{-1}.
     \end{equation}
With the Wronskian relation  $r^2[j_l(\kappa^2, r) n'_{l}(\kappa^2, r) - n_{l}(\kappa^2, r)j'_l(\kappa^2, r)] =  1$, we finally write the relation of $S^a$ and $M^a$ matrices as
    \begin{equation}
         S^a(\kappa^2) = \frac{1}{aj(\kappa^2, a)} M^a(\kappa^2) +  \frac{aj'(\kappa^2, a)}{j(\kappa^2, a)}.  
    \end{equation}

\nocite{*}

\bibliography{PRB-ASSW.bib}

\end{document}